\documentclass[a4paper,11pt]{article}
\usepackage[
  left=2.2cm,
  right=2.2cm,
  top=2cm,
  bottom=2.5cm,
  footskip=35pt 
]{geometry}

\renewcommand{\arraystretch}{1.2} 
\usepackage{caption}
\DeclareCaptionFormat{myformat}{%
  #1#2#3\par\vspace{4pt}\centerline{\rule{0.77\linewidth}{0.4pt}}
}
\usepackage[utf8]{inputenc}
\usepackage{multirow}
\usepackage[pdftex]{graphicx}
\usepackage{amssymb}
\usepackage{amsmath}
\usepackage{cite}
\usepackage{braket}
\usepackage{slashed}
\usepackage{longtable}
\usepackage{booktabs}
\usepackage{array}
\usepackage[normalem]{ulem}
\usepackage{braket}

\usepackage[dvipsnames]{xcolor}
\usepackage{xspace}

\usepackage{subcaption}
\usepackage{chngcntr}
\counterwithout{subfigure}{figure}

\usepackage[]{hyperref}
\allowdisplaybreaks
\numberwithin{equation}{section}

\newcommand{\appsection}[1]{%
  \refstepcounter{APP}%
  \section*{Appendix \theAPP: #1}%
  \addcontentsline{toc}{section}{\theAPP~#1}%
  \setcounter{equation}{0}%
}

\newcommand{\refapp}[1]{\hyperref[app:#1]{Appendix~\ref*{app:#1}}}
\newcommand{\reffig}[1]{\hyperref[fig:#1]{Figure~\ref*{fig:#1}}}
\newcommand{\reffigs}[2]{\hyperref[fig:#1]{Figure~\ref*{fig:#1}}~and~\hyperref[fig:#2]{Figure~\ref*{fig:#2}}}
\newcommand{\refeq}[1]{\hyperref[eq:#1]{Eq.~(\ref*{eq:#1})}}
\newcommand{\refeqs}[2]{\hyperref[eq:#1]{Eqs.~(\ref*{eq:#1})-(\ref*{eq:#2})}}
\newcommand{\refeqa}[2]{\hyperref[eq:#1]{Eqs.~(\ref*{eq:#1})}~and~\hyperref[eq:#2]{(\ref*{eq:#2})}}
\newcommand{\refsec}[1]{\hyperref[sec:#1]{Section~\ref*{sec:#1}}}
\newcommand{\refsubsec}[1]{\hyperref[sec:#1]{Subsection~\ref*{sec:#1}}}
\newcommand{\reftab}[1]{\hyperref[tab:#1]{Table~\ref*{tab:#1}}}

\newcommand{\C}{\mathcal{C}}

\newcommand{\cO}{\mathcal{O}}

\newcommand{\had}{\text{had}}
\newcommand{\OPE}{\text{OPE}}

\renewcommand{\Im}{\text{Im\,}}
\newcommand{\GeV}{\text{GeV}}
\newcommand{\MeV}{\text{MeV}}

\newcommand{\MSbar}{$\overline{\text{MS}}$\xspace}

\newcommand{\order}[1]{\mathcal{O}\left(#1\right)}

\newcounter{TODO}

\newcounter{TODOd}

\makeatletter
\DeclareRobustCommand{\cev}[1]{%
  {\mathpalette\do@cev{#1}}%
}
\newcommand{\do@cev}[2]{%
  \vbox{\offinterlineskip
    \sbox\z@{$\m@th#1 x$}%
    \ialign{##\cr
      \hidewidth\reflectbox{$\m@th#1\vec{}\mkern4mu$}\hidewidth\cr
      \noalign{\kern-\ht\z@}
      $\m@th#1#2$\cr
    }%
  }%
}
\makeatother

\begin{document}

\vspace*{10mm}
\begin{center}
\fontsize{15}{20}\selectfont
\bf
\boldmath
    Decay constants of the $B_c$ mesons with vector and tensor currents
\end{center}

\vspace{-2mm}

\begin{center}
{Eduard Costa i Reina, Nico Gubernari, Zachary Wüthrich}\\[5mm]
{\it\small
Helmholtz-Institut f\"ur Strahlen- und Kernphysik, Universit\"at Bonn, 53115 Bonn, Germany
\\[2mm]
E-mail:{\textnormal{ 
\texttt{eduardcostareina@gmail.com}, 
\texttt{nicogubernari@gmail.com}, 
\texttt{zjb.wuthrich@gmail.com}}}
}
\end{center}

\vspace{20mm}
\begin{abstract}\noindent
\vspace{-5mm}

\noindent
We calculate the decay constants of the lowest-lying $B_c$ mesons in the spin-parity channels $J^P = 0^-,\,0^+,\,1^-,\,1^+$, commonly referred to as the $B_c,\, B_{c0}^*,\, B_c^*,\, B_{c1}$ mesons, respectively.
Within the framework of QCD sum rules, we consider the decay constants associated with both the (axial-)vector and (axial-)tensor interpolating currents.
The decay constants with (axial-)vector currents have already been studied in the literature.
We refine previous QCD sum rule results by including higher-order power corrections and performing a comprehensive uncertainty analysis.
Furthermore, we provide the first determination of the (axial-)tensor decay constants of the $B_c^*$ and $B_{c1}$ mesons.
These new results will not only improve theoretical predictions for purely leptonic $B_c$ decays, but also strengthen unitarity constraints on $b \to c$ form factors, thereby improving the precision of predictions for $\bar{B} \to D^{(*)} \ell \bar\nu$ decays and the determination of $|V_{cb}|$.
\end{abstract}

\vspace*{2cm}

\tableofcontents

\newpage

\section{Introduction}
\label{sec:intro}

Semileptonic decays induced by the charged-current transition $b \to c \ell \bar\nu$ play a central role in flavour physics, primarily through the golden channels $\bar{B} \to D \ell \bar\nu$ and $\bar{B} \to D^* \ell \bar\nu$.
They provide a clean environment to probe the structure of the weak interaction, to test the universality of lepton couplings, and to determine the CKM element $|V_{cb}|$. 
In $b \to c \ell \bar\nu$ decays, the leptonic and hadronic contributions to the amplitude factorise when QED corrections are neglected, allowing for precise theoretical predictions within the Standard Model and sensitivity to possible new-physics effects.

Decays of the $B_c$ meson provide a complementary probe of the $b \to c \ell \bar\nu$ transition.
In particular, the purely leptonic decay $\bar{B}_c \to \ell \bar\nu$ is theoretically clean, as the hadronic effects are encapsulated in a single decay constant $f_{B_c}$, defined as
\begin{align}
    \bra{0} \bar{c} \gamma^\mu \gamma_5 b \ket{\bar{B}_c (q)} 
    =
    i \, q^\mu  f_{B_c}  \,.
\end{align}
The decay $\bar{B}_c \to \tau \bar\nu$ can be measured at the LHCb experiment~\cite{deJong2022,Galati:2026wrk,Galati:2026mqn}.
In fact, despite the small $B_c$ production fraction --- only approximately $\mathcal{O}(10^{-3})$ of all $b$ hadrons at the LHC --- the very large $b\bar{b}$ production cross section makes such measurements feasible.
As a result, sizeable $B_c$ samples are already available, and much larger ones are expected in future runs of the LHC. 
Fully leptonic decays of the $B_c$ meson could likewise be measured at FCC-ee~\cite{Amhis:2021cfy}.

The $B_c$ decay constants are not only important for predicting purely leptonic decay rates, but also enter into the unitarity bounds for $b\to c$ form factors.
These bounds follow from a dispersive representation of suitable two-point correlators, after inserting a complete set of hadronic states and using the positivity of the corresponding spectral functions.
Schematically, the bound for the spin--parity channel $0^-$ can be written as (see, e.g., Refs.~\cite{Boyd:1997kz,Caprini:1997mu,Gubernari:2026sqc})
\begin{align}
    \chi_{0^-} > 
    \left|f_{B_c}\right|^2 
    + 
    \int\limits_{(m_B+m_{D^*})^2}^\infty ds \, 
    \left|A_0^{BD^*}(s)\, W(s)\right|^2  \,,
\end{align}
where $\chi_{0^-}$ is a calculable quantity, $A_0^{BD^*}$ is a $\bar{B}\to D^*$ form factor, and $W$ is a known weight function.
Since $f_{B_c}$ contributes directly to the saturation of the unitarity bound, an accurate determination of $f_{B_c}$ tightens the allowed range of the $A_0^{BD^*}$ normalisation and shape.
Similar unitarity bounds for the $0^+$, $1^-$, and $1^+$ spin--parity channels were first established nearly thirty years ago~\cite{Boyd:1997kz,Caprini:1997mu}.
Moreover, unitarity bounds associated with the tensor currents
$\bar{c}\,\sigma^{\mu\alpha}(\gamma_5)\,q_\alpha b$, labelled as $1_T^-$ and $1_T^+$, have recently been derived~\cite{Bordone:2025jur}.
An accurate determination of the decay constants of the lowest-lying $\bar{c}b$ states in each of these spin--parity channels is therefore essential to further strengthen these bounds.
These states and their masses are listed in \reftab{spectr}.
Stronger bounds will enhance the overall precision of the $B \to D^{(*)}$ (and $B_s \to D_s^{(*)}$) form factors.
This, in turn, will lead to more precise 
phenomenological predictions for $\bar{B}\to D^{(*)}\ell\bar{\nu}$ decays and, consequently, to more accurate determinations of $|V_{cb}|$ from exclusive channels~\cite{Caprini:1997mu,Martinelli:2023fwm,Bordone:2024weh,Bordone:2025jur}.
\bigskip

The calculation of the decay constants is a challenging non-perturbative problem in QCD, since these quantities encode long-distance strong-interaction dynamics that cannot be accessed within perturbation theory alone.
There are two main theoretical frameworks currently used to calculate these decay constants: lattice QCD and QCD sum rules.

Only the decay constants of the $B_c$ and $B_c^*$ mesons with (axial-)vector currents have been determined using lattice QCD~\cite{Colquhoun:2015oha,Becirevic:2018qlo,Cai:2026xja}.
This limitation arises because lattice QCD calculations are computationally demanding, and the treatment of unstable QCD states remains particularly challenging within this framework.
These decay constants have also been determined within the non-relativistic QCD factorization framework; see, e.g., Refs.~\cite{Tao:2022qxa,Tao:2023mtw}.

All decay constants of the mesons listed in \reftab{spectr} associated with (axial-)vector currents have been calculated previously using QCD sum rules~\cite{Colangelo:1992cx,Kiselev:2000pp,Kiselev:2002vz,Kiselev:2003uk,Narison:2019tym,Wang:2012kw,Aliev:2019wcm,Narison:2020wql,Wang:2024fwc}.
In this article, we reassess and refine these determinations.
In particular, we extend the relevant operator product expansions (OPEs) entering the QCD sum rules by including corrections up to dimension six together with next-to-leading-order (NLO) $\alpha_s$ corrections to the leading-power term, i.e.\ the perturbative contribution.
Even more importantly, we perform a careful uncertainty analysis by varying all relevant input parameters, as well as the Borel parameter and the renormalisation scale, over appropriate ranges in order to obtain realistic uncertainties and avoid underestimated errors.
Most notably, we present for the \emph{first} time QCD sum rule determinations of the $B_c^*$ and $B_{c1}$ meson decay constants associated with \mbox{(axial-)tensor} currents.
Although (axial-)tensor currents do not contribute within the Standard Model, they appear in many realistic extensions of it.
\\

\begin{table}[t]
    \centering
	\begin{tabular}{cccc}
        \toprule
	  Meson & $J^P$ & Mass [\GeV] & Ref. \\
		\midrule
		$B_c$       &  $0^-$ & $6.2745(3)$ & \cite{PDG:2024cfk}\\
		$B_{c0}^*$  &  $0^+$ & $6.707(16)$ & \cite{Dowdall:2012ab}\\
		$B_c^*$     &  $1^-$ & $6.3390(20)$ & \cite{ATLAS:2026ubk}\\
		$B_{c1}$    &  $1^+$ & $6.742(16)$ & \cite{Dowdall:2012ab}\\
		\bottomrule
	\end{tabular}
	\caption{
        Mass, spin ($J$), and parity ($P$) of the $B_c$ mesons considered in this article.
        For the central values and uncertainties of the masses, we use experimental measurements whenever they are available; otherwise, we rely on lattice QCD determinations.
	\label{tab:spectr}
	}
\end{table}

The remainder of this article is organised as follows.
In \refsec{calc} we define the decay constants and the relevant two-point correlators, derive their hadronic and OPE representations, and construct the QCD sum rules used in the analysis.
\refsec{results} contains the numerical extraction of the decay constants, the stability and uncertainty analysis, the comparison with previous determinations, and the discussion of the resulting one-particle contributions to the $b\to c$ unitarity bounds.
We summarise our findings in \refsec{summary}.
The derivation of the vacuum-averaged gluon-condensate decompositions is given in \refapp{Condensates}, while further details on the numerical stability checks are collected in \refapp{Numerics}.

\section{QCD sum rule calculation}
\label{sec:calc}

Decay constants are QCD parameters that describe how strongly a hadron couples to a given quark current.
Mathematically, they are defined through the hadron-to-vacuum matrix elements of the corresponding quark current.
In this work, we consider the following six decay constants:
\begin{equation}
\begin{aligned}
    \label{eq:def-dec-const}
    \bra{0} J_V^\mu \ket{\bar{B}_{c0}^*(q)} 
    & =
    i \, q^\mu f_{B_{c0}^*} \,,
    &
    \bra{0} J_{A}^\mu \ket{\bar{B}_c(q)} 
    & =
    i \, q^\mu f_{B_c} \,,
    \\
    \bra{0} J_V^\mu \ket{\bar{B}_c^*(q,\eta)} 
    & =
    m_{B_c^*} \eta^\mu f_{B_c^*} \,,
    &
    \bra{0} J_{A}^\mu \ket{\bar{B}_{c1}(q,\eta)} 
    & =
    m_{B_{c1}} \eta^\mu f_{B_{c1}} \,,
    \\
    \bra{0} J_T^\mu \ket{\bar{B}_c^*(q,\eta)} 
    & =
    i \, m_{B_c^*}^2 \eta^\mu f_{B_c^*}^T \,,
    &
    \bra{0} J_{AT}^\mu \ket{\bar{B}_{c1}(q,\eta)} 
    & =
    i \, m_{B_{c1}}^2 \eta^\mu f_{B_{c1}}^T \,,
\end{aligned}
\end{equation} 
where $q$ is the momentum injected by the current and $\eta$ is the spin-one mesons' polarisation vector.
The quark currents are defined as
\begin{equation}
\begin{aligned}
    \label{eq:JGamma}
    J_V^\mu & = \bar{c}\, \gamma^\mu b    \,,&\qquad
    J_A^\mu & = \bar{c}\, \gamma^\mu \gamma_5 b      \,,\\
    J_T^\mu & = \bar{c}\, \sigma^{\mu \alpha}q_\alpha  b   \,,&
    J_{AT}^\mu & = \bar{c}\, \sigma^{\mu \alpha}q_\alpha \gamma_5  b
    \,.
\end{aligned}
\end{equation}

We calculate the decay constants in~\refeq{def-dec-const} following the standard procedure of Ref.~\cite{Shifman:1978bx} for deriving QCD sum rules, see also Refs.~\cite{Reinders:1984sr,Colangelo:2000dp} for reviews.
To this end we define the following correlators
\begin{align}
    \label{eq:corr}
    \Pi_\Gamma^{\mu\nu}(q)
    & \equiv 
    i\! \int\! d^4x\, e^{iq\cdot x} 
        \bra{0} 
            T \left\{J_\Gamma^{\mu}(x) \, J_\Gamma^{\dagger,\nu}(0)\right\} 
        \ket{0}
    =
    \left(\frac{q^\mu q^\nu}{q^2} - g^{\mu\nu}\right)
    \Pi_\Gamma^1(q^2)
    +
    \frac{q^\mu q^\nu}{q^2} \,\Pi_\Gamma^0(q^2)
    \,,
\end{align}
where $\Gamma = V,\,A,\,T,\,AT$.
Note that $\Pi_T^0=\Pi_{AT}^0=0$ since $\sigma^{\mu \alpha}q_\alpha q_\mu = 0$.
The correlators $\Pi_\Gamma^J$ satisfy a dispersion relation (see, e.g., Ref.~\cite{Colangelo:2000dp,Gubernari:2026sqc})
\begin{align}
    \label{eq:disprel}
    \Pi_\Gamma^J(q^2) 
    & =
    \frac{1}{\pi} \int\limits_{0}^\infty d s \, \frac{\Im\Pi_\Gamma^J(s)}{s - q^2 - i\varepsilon} 
    \,.
\end{align}
In the remainder of this article, the factor $i\varepsilon$ is omitted for brevity.
We note that $\Im\Pi_\Gamma^J(s)$ vanishes below the lowest production threshold in the corresponding channel, as shown explicitly below.
Consequently, although the lower integration limit in Eq.~(2.4) is written as zero, the region below this threshold does not contribute.
Moreover, the correlators $\Pi_\Gamma^J$ generally exhibit ultraviolet divergences.
To handle these divergences, one can either perform subtractions in the dispersion relation or apply a Borel transform.
In our analysis, we adopt the Borel transform, which is particularly advantageous in the framework of QCD sum rules, as it exponentially suppresses contributions from higher excited states and continuum thresholds~\cite{Colangelo:2000dp}.
Nonetheless, we also verify the stability of our results by performing the subtraction procedure, as discussed below.

After applying the Borel transform to \refeq{disprel}, the dispersion relation becomes
\begin{align}
    \label{eq:dispborel}
    \Pi_\Gamma^J(M^2)
    &= 
    \frac{1}{\pi} 
    \int\limits_0^\infty ds \, e^{-s/M^2} \, \Im\Pi_\Gamma^J(s)
    \,,
\end{align}
where $M^2$ denotes the Borel parameter.

The QCD sum rules are obtained by matching the hadronic representation of the correlators $\Pi_\Gamma^J$, presented in \refsubsec{hadrep}, to their corresponding OPE calculation, discussed in \refsubsec{OPEcal}.

\subsection{Hadronic representation of the correlators}
\label{sec:hadrep}

The imaginary part of $\Pi_\Gamma^J$ can be written in terms of hadronic quantities using unitarity (see, e.g., Refs.~\cite{Colangelo:2000dp,Khodjamirian:2020btr}).
For $\Pi_A^0$, one finds
\begin{align}
    \label{eq:hadA0}
    \frac{1}{\pi} \Im\Pi_A^0(s) \Big|_\had
    =
    m_{B_c}^2 f_{B_c}^2 
    \delta\!\left(s - m_{B_c}^2\right) + 
    \rho_A^0(s) \, \theta\!\left(s - s_A^0 \right)
    \,,
\end{align}
and analogous formulae hold for $\Pi_V^0$, $\Pi_V^1$, and $\Pi_A^1$.
For $\Pi_T^1$,  one finds
\begin{align}
    \label{eq:hadT}
    \frac{1}{\pi} \Im\Pi_T^1(s) \Big|_\had
    =
    m_{B_c^*}^4 \left( f_{B_c^*}^T \right)^2 
    \delta\!\left(s - m_{B_c^*}^2\right) + 
    \rho_T^1(s) \, \theta\!\left(s - s_T^1 \right)
    \,,
\end{align}
and an analogous formula holds for $\Pi_{AT}^1$.
Here, $\rho_\Gamma^J$ denotes the spectral density function, which includes the contributions from the continuum and excited states above the hadronic threshold~$s_\Gamma^J$.

Note that the widths of the mesons under consideration are not known, and all states are therefore treated as narrow. 
This is a very good approximation for the $B_c$ and $B_c^*$ mesons, which are QCD-stable. 
Nevertheless, the excited states $B_{c0}^*$ and $B_{c1}$ may exhibit sizeable decay widths, although they remain below the $BD$ threshold.
Once the widths of these states are determined, \refeqs{hadA0}{hadT} can be straightforwardly modified to account for these effects (see, e.g., Refs.~\cite{Gelhausen:2014jea,Gubernari:2022hrq,Gubernari:2023rfu}).

A further subtlety arises in the scalar and axial-vector channels.
In these channels, the spectral functions may also contain contributions from multi-hadron states, such as $B_c\pi\pi$ and $B_c^*\pi\pi$.
The corresponding production thresholds lie below the masses of the $B_{c0}^*$ and $B_{c1}$ mesons.
In the present analysis, we assume that --- in the low-lying part of the spectrum --- the spectral density is dominated by the $B_{c0}^*$ and $B_{c1}$ mesons.
The corresponding multi-hadron contributions are expected to be subleading, as is typical for excited $B_c$ states below the $BD$ threshold~\cite{Ortega:2020uvc,Martin-Gonzalez:2022qwd}.

\subsection{OPE of the correlators}
\label{sec:OPEcal}

The correlators $\Pi_\Gamma^J$ can be calculated using a local operator product expansion (OPE) in the kinematic regime where \hbox{$q^2 \ll (m_b + m_c)^2$}~\cite{Colangelo:2000dp}.
This OPE can be organised as a series of local, vacuum-averaged operators of increasing mass dimension $d$, each multiplied by its corresponding Wilson coefficient.
After performing the Borel transform, it can be written as 
\begin{align}
    \Pi_\Gamma^J(M^2)\Big|_{\OPE}
    =
    \sum_{d=0}^\infty
    \C_{\Gamma,d}^J(M^2)\,
    \langle\cO_d\rangle
    \,,
\end{align}
where $\langle\cO_d\rangle := \bra{0}\cO_d\ket{0}$.
The Wilson coefficients $\C_{\Gamma,d}^J(M^2)$ depend on the choice of interpolating current and describe the short-distance contributions to the correlator.
In contrast, the vacuum expectation values $\bra{0}\cO_d\ket{0}$ encode the long-distance, genuinely non-perturbative effects.
The former are calculable in perturbation theory, whereas the latter are parametrised in terms of universal vacuum condensates.

We include contributions up to $d=6$.
The impact of higher-dimensional contribution is strongly suppressed and can safely be neglected within the precision of the present analysis~\cite{Colangelo:2000dp,Khodjamirian:2020btr}.
In addition, in purely heavy--quark systems, condensates involving heavy--quark fields do not appear directly~\cite{Reinders:1984sr,Bagan:1994dy}. 
Therefore, the OPE for $\Pi_\Gamma^J$ contains three types of contributions (see, e.g., Refs.~\cite{Nikolaev:1982rq,Colangelo:2000dp}):
\begin{align}
    \label{eq:OPEexpl}
    \Pi_\Gamma^J(M^2) \Big|_\OPE 
    \simeq
    \C_{\Gamma,0}^J (M^2)
    +
    \C_{\Gamma,4}^J (M^2) \langle \cO_4 \rangle 
    +
    \C_{\Gamma,6}^J (M^2) \langle \cO_6 \rangle 
    \,.
\end{align}
Here we have used the fact that the identity is the only operator of dimension zero,
\begin{align}
    \langle \cO_0 \rangle := \langle 0 | \mathbb{I} | 0 \rangle \equiv 1
    \,,
\end{align}
while the two-gluon condensate is the only operator at dimension four,
\begin{align}
    \langle \cO_4 \rangle &:=  \bra{0} \frac{\alpha_s}{\pi} G^a_{\mu\nu} G^a_{\mu\nu} \ket{0} \,.
\end{align}
At dimension six, by contrast, there are two linearly independent operators, such that
\begin{align}
    \label{eq:C6}
    \C_{\Gamma,6}^J(M^2) \langle \cO_6 \rangle 
    &:= 
    \C_{\Gamma,GGG}^J (M^2) \bra{0} g_s^3 f^{abc} G^a_{\mu\nu} G^b_{\nu\lambda} G^c_{\lambda\mu} \ket{0} + \C_{\Gamma,jj}^J (M^2) \bra{0} g_s^4 j^a_\mu j^a_\mu \ket{0} \, ,
\end{align}
where light-quark currents are defined by
\begin{align}
    j^a_\mu = \sum_{\psi = u,d,s} \bar{\psi} \gamma_\mu t^a \psi
    \,.
\end{align}
The matrices $t^a$ denote the generators of $SU(3)_C$ in the fundamental representation and satisfy $t^a=\lambda^a/2$, where $\lambda^a$ are the Gell-Mann matrices.
As we explain below (cf. \refeq{gluonEOM}), the light--quark condensates arise indirectly from dimension six two gluon condensates with two covariant derivatives.

\begin{figure}[t!]
    \centering
    \begin{subfigure}[b]{0.2\textwidth}
        \centering
        \includegraphics[width=\textwidth]{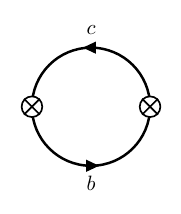}
        \subcaption{}\label{fig:LPa}
    \end{subfigure}
    \begin{subfigure}[b]{0.2\textwidth}
        \centering
        \includegraphics[width=\textwidth]{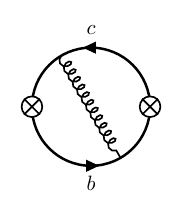}
        \subcaption{}\label{fig:LPb}
    \end{subfigure}
    \begin{subfigure}[b]{0.2\textwidth}
        \centering
        \includegraphics[width=\textwidth]{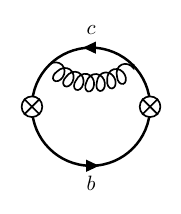}
        \subcaption{}\label{fig:LPc}
    \end{subfigure}
    \begin{subfigure}[b]{0.2\textwidth}
        \centering
        \includegraphics[width=\textwidth]{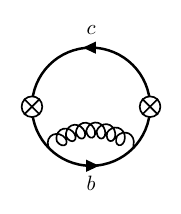}
        \subcaption{}\label{fig:LPd}
    \end{subfigure}
    \caption{
        Perturbative diagrams entering the leading-power Wilson coefficient $\C_{\Gamma,0}^J$.
        Diagram~\subref{fig:LPa} gives the LO contribution, while diagrams~\subref{fig:LPb}--\subref{fig:LPd} give the NLO corrections in $\alpha_s$.
        The crossed circles denote the insertion points of the interpolating currents~$J_\Gamma^\mu$.
    }
    \label{fig:LP}
\end{figure}

The leading-power term in \refeq{OPEexpl} is obtained entirely within perturbation theory and is therefore referred to as the \emph{perturbative contribution}.
In our analysis, we include the NLO corrections in $\alpha_s$ to $\C_{\Gamma,0}^J$~\cite{Djouadi:1993ss,Generet:2025hsv}.
The corresponding diagrams are shown in \reffig{LP}.
For later convenience, we write the perturbative contribution in a dispersive form (cf. \refeq{dispborel}):
\begin{equation}
\label{eq:pertdisp}
\begin{aligned}
    \C_{\Gamma,0}^J (M^2)
    & =
    \frac{1}{\pi} \int\limits_{0}^\infty d s 
    \, e^{-s/M^2} \, 
    \Im\C_{\Gamma,0}^J (s)
    \\
    & =
    \frac{1}{\pi} \int\limits_{0}^\infty d s 
    \, e^{-s/M^2} \, 
    \left(
        \Im\C_{\Gamma,0}^J (s) \Big|_\text{LO}
        +
        \Im\C_{\Gamma,0}^J (s) \Big|_\text{NLO}
    \right) + \dots \,. 
\end{aligned}
\end{equation}
We take the expressions for the LO and NLO imaginary parts of $\C_{\Gamma,0}^J$ from the ancillary files provided with the arXiv version of Ref.~\cite{Generet:2025hsv}.

The next-to-leading-power and next-to-next-to-leading-power terms in the OPE correspond to the dimension-four and dimension-six contributions, respectively.
For vector currents, the Wilson coefficients of the dimension-four operator are available in the literature; see, e.g., Refs.~\cite{Reinders:1984sr,Colangelo:1992cx}.
We have independently verified these results and find complete agreement.
The Wilson coefficients of the dimension-six operator were calculated in the limit $m_c = m_b$ in Refs.~\cite{Nikolaev:1981ff,Nikolaev:1982ra,Nikolaev:1982rq,Novikov:1983gd}.
We have independently checked these expressions as well and again find complete agreement.
In addition, we derive new analytical expressions valid for $m_c \neq m_b$.\footnote{
    Ref.~\cite{Bagan:1994dy} presents results for the pseudoscalar current in the case $m_c \neq m_b$.
    We noted that the expressions given in Eq.~(117) of that reference contain a few typographical errors.
    More precisely, $m$ must be replaced by $m_c$ in the coefficient of $t^2$, the polynomial multiplying $t$ must contain $3m_c^6$ rather than $m_c^6$, and the factor multiplying the $C$ coefficient must read $(t-\Sigma^2)^3$. 
}
For tensor currents, the Wilson coefficients of the dimension-four and dimension-six operators are derived here for the first time.
This constitutes one of the main new results of the present work.
\\

The calculation of these subleading Wilson coefficients is non-trivial and we describe it in what follows.
Throughout this work, we use the conventions
\begin{align}
\begin{aligned}
    g_{\mu\nu}
    &= {\rm diag} (+,-,-,-)
    \,,&
    \sigma_{\mu\nu}
    &=
    \frac{i}{2}
    [\gamma_\mu,\gamma_\nu]
    \,,&
    \gamma_5 &= i\,\gamma^0 \gamma^1 \gamma^2 \gamma^3 \, ,
    \\
    (D_\alpha)^{ab} &= \delta^{ab}\partial_\alpha + g_s f^{abc} A^b_\alpha,
    &
    [ D_\alpha, D_\beta ]^{bc} &=
    -g_s\,f^{abc} G_{\alpha \beta}^a ,&
    \epsilon^{0123}
    &=
    +1
    \,,
\end{aligned}
\end{align}
where $f^{abc}$ denote the antisymmetric $SU(3)_C$ structure constants, and the covariant derivatives act in the adjoint colour representation.
In the framework of the local OPE, the correlator in \refeq{corr} can be written as
\begin{equation}
\label{eq:corrposition}
    \Pi_\Gamma^{\mu\nu}(q)=-i\int d^4x \, e^{iqx}\bra{0}\text{Tr}[\Gamma^\mu  S_b(x,0) \Gamma^{\dagger,\nu} S_c(0,x)]\ket{0}\, .
\end{equation}
Here, we define $S_\psi(x,y):=\bra{0}T\{\psi(x)\bar{\psi}(y)\}\ket{0}$, while the trace is understood to run over the fundamental colour and Dirac spaces.
The non-trivial part of the calculation that gives rise to the gluon-condensate contributions is the treatment of the quark propagator in the background gluon field.
In practice, this amounts to expanding the propagator in the background field associated with the vacuum gluons.
Following Ref.~\cite{Novikov:1983gd}, we express the quark propagator in the background vacuum gluon field as
\begin{equation} \label{eq:modprop}
\begin{aligned}
    S_\psi(x,y)
    &=
    S_\psi^{(0)}(x-y)+g_s\int d^4z\,S_\psi^{(0)}(x-z)\, i\slashed A(z)\,S_\psi^{(0)}(z-y)\\
    &+ g_s^2\int d^4z'd^4z\,S_\psi^{(0)}(x-z')\, i\slashed A(z')\,S_\psi^{(0)}(z'-z)\, i\slashed A(z)\,S_\psi^{(0)}(z-y)\, + \dots\,,
\end{aligned}
\end{equation}
where $S_\psi^{(0)}(x-y)$ denotes the free quark propagator, and $A_\mu(x)=t^aA_\mu^a(x)$ is the background vacuum gluon field.
For later convenience, we introduce the following shorthand notation for covariant derivatives acting on the gluon field-strength tensor:
\begin{equation}
    G_{\mu\nu;\alpha_1\cdots\alpha_n}^a:=(D_{\alpha_1})^{aa_1}(D_{\alpha_2})^{a_1a_2}\dots (D_{\alpha_n})^{a_{n-1}a_n}G_{\mu\nu}^{a_n}\,.
\label{eq:GDnotation}
\end{equation}
Using this notation, the expansion of the gluon field in the Fock--Schwinger gauge~\cite{Schwinger:1951nm,Novikov:1983gd} reads
\begin{equation}
    A_\mu^a(x,z_0) = \sum_{n=0}^{N}\frac{1}{n!(n+2)} (x-z_0)^\nu (x-z_0)^{\alpha_1} \cdots (x-z_0)^{\alpha_n} G_{\nu\mu;\alpha_1\cdots\alpha_n}^a\ ,
\label{eq:gluonFSgauge}
\end{equation}
where $z_0$ may be viewed as a gauge parameter, which must drop out of all final results.
Equation~\eqref{eq:gluonFSgauge} is equivalent to Eq.~(3.12) of Ref.~\cite{Nikolaev:1982rq} and Eq.~(2.7) of Ref.~\cite{Novikov:1983gd}.
For the calculation of the dimension-four and dimension-six contributions, it is sufficient to retain terms up to $N=2$ in the expansion, i.e.\ terms containing at most two covariant derivatives.
Accordingly, in \refeq{corrposition} we keep only those composite operators whose mass dimension does not exceed six.

\begin{figure}[p]
    \setcounter{subfigure}{4}
    \centering
    \begin{subfigure}[b]{0.2\textwidth}
        \centering
        \includegraphics[width=\textwidth]{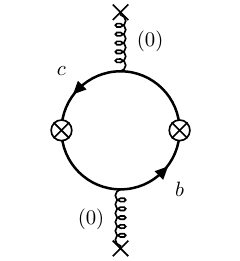}
        \subcaption{}\label{fig:GGconda}
    \end{subfigure}
    \begin{subfigure}[b]{0.2\textwidth}
        \centering
        \includegraphics[width=\textwidth]{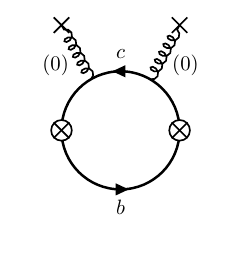}
        \subcaption{}\label{fig:GGcondb}
    \end{subfigure}
    \begin{subfigure}[b]{0.2\textwidth}
        \centering
        \includegraphics[width=\textwidth]{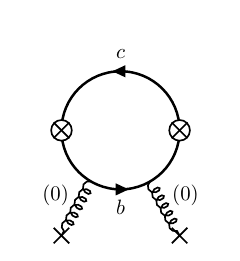}
        \subcaption{}\label{fig:GGcondc}
    \end{subfigure}
    \caption{ 
        Background-field diagrams contributing to the dimension-four terms in the local OPE.
        The crossed circles denote the insertion points of the interpolating currents~$J_\Gamma^\mu$.
        The wavy lines ending in crosses denote background-gluon insertions in the external-field expansion of the heavy--quark propagators.
        \label{fig:GGcond}
    }
    \vspace{1cm}
    \setcounter{subfigure}{7}
    \begin{subfigure}[b]{0.18\textwidth}
        \centering
        \includegraphics[width=\textwidth]{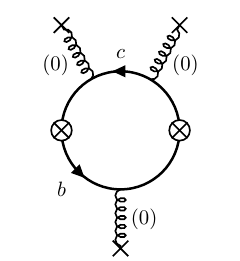}
        \subcaption{}\label{fig:GGGconda}
    \end{subfigure}
    \begin{subfigure}[b]{0.18\textwidth}
        \centering
        \includegraphics[width=\textwidth]{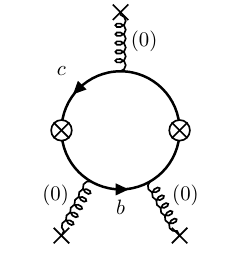}
        \subcaption{}\label{fig:GGGcondb}
    \end{subfigure}
    \begin{subfigure}[b]{0.18\textwidth}
        \centering
        \includegraphics[width=\textwidth]{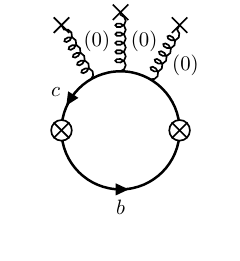}
        \subcaption{}\label{fig:GGGcondc}
    \end{subfigure}
    \begin{subfigure}[b]{0.18\textwidth}
        \centering
        \includegraphics[width=\textwidth]{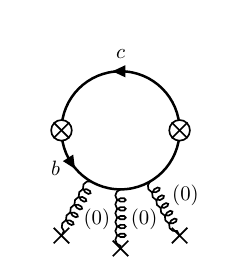}
        \subcaption{}\label{fig:GGGcondd}
    \end{subfigure}
    \begin{subfigure}[b]{0.18\textwidth}
        \centering
        \includegraphics[width=\textwidth]{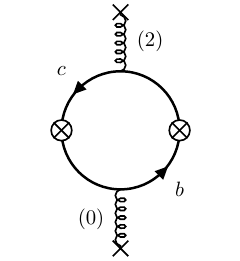}
        \subcaption{}\label{fig:GGGconde}
    \end{subfigure}
    \\[0.3cm]
    \begin{subfigure}[b]{0.18\textwidth}
        \centering
        \includegraphics[width=\textwidth]{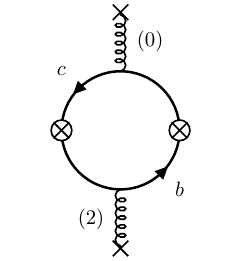}
        \subcaption{}\label{fig:GGGcondf}
    \end{subfigure}
    \begin{subfigure}[b]{0.18\textwidth}
        \centering
        \includegraphics[width=\textwidth]{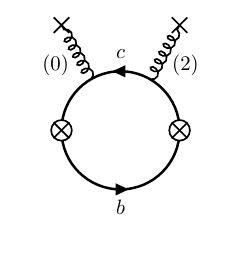}
        \subcaption{}\label{fig:GGGcondg}
    \end{subfigure}
    \begin{subfigure}[b]{0.18\textwidth}
        \centering
        \includegraphics[width=\textwidth]{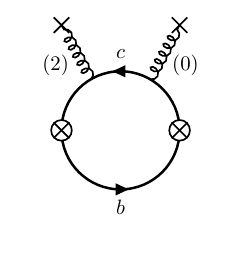}
        \subcaption{}\label{fig:GGGcondh}
    \end{subfigure}
    \begin{subfigure}[b]{0.18\textwidth}
        \centering
        \includegraphics[width=\textwidth]{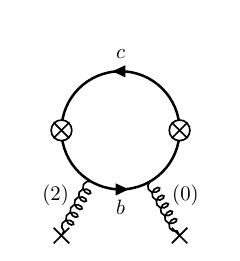}
        \subcaption{}\label{fig:GGGcondi}
    \end{subfigure}
    \begin{subfigure}[b]{0.18\textwidth}
        \centering
        \includegraphics[width=\textwidth]{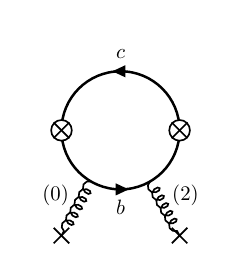}
        \subcaption{}\label{fig:GGGcondj}
    \end{subfigure}
    \\[0.3cm]
    \begin{subfigure}[b]{0.18\textwidth}
        \centering
        \includegraphics[width=\textwidth]{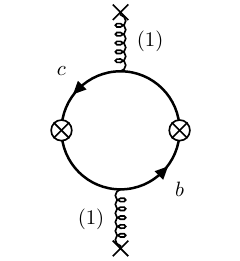}
        \subcaption{}\label{fig:GGGcondk}
    \end{subfigure}
    \begin{subfigure}[b]{0.18\textwidth}
        \centering
        \includegraphics[width=\textwidth]{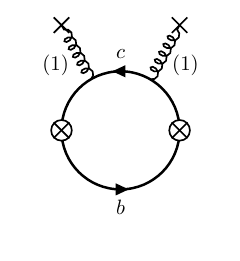}
        \subcaption{}\label{fig:GGGcondl}
    \end{subfigure}
    \begin{subfigure}[b]{0.18\textwidth}
        \centering
        \includegraphics[width=\textwidth]{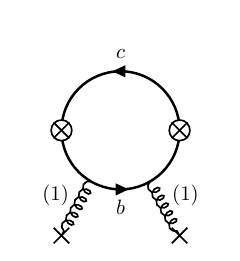}
        \subcaption{}\label{fig:GGGcondm}
    \end{subfigure}
    \captionsetup{format=plain} 
    \caption{
        Background-field diagrams contributing to the dimension-six terms in the local OPE.
        The crossed circles denote the insertion points of the interpolating currents~$J_\Gamma^\mu$.
        The wavy lines ending in crosses denote background-gluon insertions in the external-field expansion of the heavy--quark propagators.
    \label{fig:GGGcond}
    }
\end{figure}

Since the heavy--quark propagators are most conveniently handled in momentum space, we perform the calculation in that representation.
To this end, we use the Fourier-transform identities
\begin{equation} \label{eq:propertiespos}
\begin{aligned}
    f(k)
    &=  
    \int d^4x e^{ikx} f(x)  \ ,\\
    \int d^4x e^{ikx} 
    &=
    (2\pi)^4 \delta^{(4)}(k)\ ,\\
    \int d^4x x^{\alpha_1}\cdots x^{\alpha_n} e^{ik\cdot x}
    &= 
    (2\pi)^4 (-i)^n \frac{\partial^n}{\partial k_{\alpha_1} \cdots \partial k_{\alpha_n}}  \delta^{(4)}(k)\ ,
\end{aligned}
\end{equation}
which convert powers of $x$ in coordinate space into derivatives acting on momentum-space delta functions.
The resulting expressions can then be evaluated by repeatedly using
\begin{equation} \label{eq:propertiesmom}
\begin{aligned}
\int d^4k f(k)  \delta^{(4)}(k) 
    &=  
    f(0)\ ,\\
    \int d^4k f(k) \Big[ \frac{\partial^n}{\partial k_{\alpha_1} \cdots \partial k_{\alpha_n}}\,\delta^{(4)}(k)\Big]
    &= 
    (-1)^n \left. \frac{\partial^n}{\partial k_{\alpha_1} \cdots \partial k_{\alpha_n}}  f(k) \right|_{k=0}\ .
\end{aligned}
\end{equation}
Taking into account all relevant combinations of $n$-gluon insertions on the quark propagators, together with the truncation of the gluon field-strength expansion at order $N$, we obtain the complete set of Feynman diagrams shown in \reffig{GGcond} and \reffig{GGGcond} for the dimension-four and dimension-six contributions, respectively.
In these figures, the label $(r)$ on a gluon insertion implies taking only the relevant term proportional to $G_{\nu \mu;\mu_1 \cdots \mu_r}$ that contains $r$ covariant derivatives.
Diagrams V,VI, and VII correspond to topologies with two gluon insertions contributing to the dimension-four term, while diagrams VIII-XX correspond to topologies with two or three gluon insertions contributing to the dimension-six term.
In fact, a two-gluon insertion accompanied by two covariant derivatives can be re-expressed in terms of a three-gluon condensate and a light four-quark condensate, as can be seen from the relations below.

The contributions associated with the two-gluon condensate (or GG condensate) are obtained by replacing the vacuum expectation value of two gluon field-strength tensors by its Lorentz decomposition~\cite{Reinders:1984sr},
\begin{equation}
    \bra{0} G_{\alpha\beta}^a G_{\mu\nu}^a  \ket{0}=\frac{1}{12} \bra{0} G_{\gamma\delta}^a G^{a}_{\gamma\delta} \ket{0}\, (g_{\alpha \mu} g_{\beta \nu} - g_{\alpha \nu} g_{\beta \mu})\,.
\end{equation}
Analogously, the three-gluon condensates diagrams arise from the substitutions
\begin{align}
    \bra{0} f_{abc} G_{\mu\nu}^a G_{\alpha\beta}^b G_{\rho\sigma}^c \ket{0} 
    &= 
    \frac{1}{24} \bra{0} f_{abc} G_{\gamma\delta}^a G^{b}_{\delta\varepsilon} G_{\varepsilon\gamma}^c \ket{0}\notag
    \\*
    &\times ( g_{\mu\sigma}g_{\alpha\nu}g_{\beta\rho} 
    + g_{\mu\beta}g_{\alpha\rho}g_{\sigma\nu} 
    + g_{\alpha\sigma}g_{\mu\rho}g_{\nu\beta} 
    + g_{\rho\nu}g_{\mu\alpha}g_{\beta\sigma}\notag \\*
    &
    - g_{\mu\beta}g_{\alpha\sigma}g_{\rho\nu} 
    - g_{\mu\sigma}g_{\alpha\rho}g_{\nu\beta} 
    - g_{\alpha\nu}g_{\mu\rho}g_{\beta\sigma} 
    - g_{\beta\rho}g_{\mu\alpha}g_{\nu\sigma} )\ ,
    \label{eq:vevGGG}
    \\[3mm]
    \bra{0} G_{\mu\nu}^a  G_{\alpha\beta;\rho\sigma}^a  \ket{0} &= \frac{g_s}{48} \bra{0} f_{abc} G_{\gamma\delta}^a G^{b}_{\delta\varepsilon} G_{\varepsilon\gamma}^{c} \ket{0}\notag
    \\* 
    &\times \big[\, 2 g_{\rho \sigma} (g_{\mu \alpha} g_{\nu \beta} - g_{\mu \beta} g_{\nu \alpha})\notag\\
    &+ g_{\mu\sigma}g_{\alpha\nu}g_{\beta\rho} 
    + g_{\mu\beta}g_{\alpha\rho}g_{\sigma\nu} 
    + g_{\alpha\sigma}g_{\mu\rho}g_{\nu\beta} 
    + g_{\rho\nu}g_{\mu\alpha}g_{\beta\sigma}\notag \\*
    &
    -\, g_{\mu\beta}g_{\alpha\sigma}g_{\rho\nu} 
    - g_{\mu\sigma}g_{\alpha\rho}g_{\nu\beta} 
    - g_{\alpha\nu}g_{\mu\rho}g_{\beta\sigma} 
    - g_{\beta\rho}g_{\mu\alpha}g_{\nu\sigma} \big]\,\notag \\
    &
    + \frac{g_s^2}{72} \bra{0} j_\lambda^a j_\lambda^a \ket{0} \notag \\
    &
    \times \big[
    2 g_{\rho  \sigma } \left(g_{\alpha  \nu } g_{\beta  \mu }
    -g_{\alpha  \mu } g_{\beta  \nu}\right) \notag \\
    &
    -g_{\alpha  \sigma } g_{\beta  \nu } g_{\mu  \rho}
    +g_{\alpha  \sigma } g_{\beta  \mu } g_{\nu  \rho }
    +g_{\alpha \nu } g_{\beta  \sigma } g_{\mu  \rho}
    -g_{\alpha  \mu } g_{\beta  \sigma } g_{\nu  \rho } 
    \notag \\
    &
    -g_{\alpha  \rho } g_{\beta  \nu } g_{\mu  \sigma }
    +g_{\alpha  \nu } g_{\beta  \rho } g_{\mu  \sigma }
    +g_{\alpha \rho } g_{\beta  \mu } g_{\nu  \sigma }
    -g_{\alpha  \mu } g_{\beta  \rho } g_{\nu  \sigma} \big]\,,
    \label{eq:vevGDDG}
    \\[3mm] 
    \bra{0} G_{\mu \nu;\rho}^a G_{\alpha \beta; \sigma}^a \ket{0} 
    &= 
    - \bra{0} G_{\mu\nu}^aG_{\alpha \beta; \sigma \rho}^a\ket{0}\ ,
    \label{eq:vevDGDG}
\end{align}
where the light-quark current $j_\mu^a$ is defined via the gluon equation of motion
\begin{align} 
\label{eq:gluonEOM}
    G^a_{\nu\mu;\mu} &
    = g_s j_\nu^a 
    = g_s \sum_{\psi=u,d,s} \bar{\psi}\, \gamma_\nu \, t^a \, \psi \,.
\end{align}
We refer to the expressions in \refeqs{vevGGG}{vevDGDG} as the GGG, GDDG, and DGDG condensates, respectively.
The GGG and GDDG condensates given in \refeqa{vevGGG}{vevGDDG} can already be found in Ref.~\cite{Nikolaev:1982rq}.
By contrast, the expression for the DGDG condensate in \refeq{vevDGDG} is not given explicitly there, although it is needed for obtaining our final results.
As mentioned above, the only condensates that arise directly in the OPE are those of gluonic operators, since heavy--quark condensates do not appear explicitly.
The light-quark condensates appear only indirectly through the gluon equation of motion and simply provide a way to parametrise the GDDG and DGDG condensates.
We re-derive \refeqs{vevGGG}{vevDGDG} in \refapp{Condensates}, as we could find no other suitable proofs of these relations in the literature.
We also use the identities $\text{Tr}[t^at^b]=\delta^{ab}/2$ and $\text{Tr}[t^at^bt^c]=(d^{abc}+if^{abc})/4$.
The term proportional to the symmetric constants $d^{abc}$ does not contribute to the final result because of the QCD Furry theorem~\cite{Nikolaev:1982rq,Smolyakov:1980wq}.
\\

We now briefly describe the contents of the ancillary file \texttt{TensorCondensates.nb}.
This file contains one of the codes used in our calculation of the dimension-four and dimension-six contributions.
In the code, the correlators in \refeq{corrposition} are implemented directly in momentum space and expressed within the OPE framework as in Refs.~\cite{Novikov:1983gd,Nikolaev:1982rq}
\begin{equation} \label{eq:GenCorr}
    \begin{aligned}
        \Pi_\Gamma^{\mu \nu}(q;n,m)=-i\int \frac{d^4k}{(2\pi)^4}\bra{0}\text{Tr}\left[\Gamma^\mu \, S_b^{(n)}(k+q)\,\Gamma^{\dagger,\nu}\,\tilde{S}_c^{(m)}(k)\right]\ket{0}\, .
    \end{aligned}
\end{equation}
Starting from \refeq{modprop}, the equivalent quark propagators $S_q^{(n)}$ and $\tilde{S}_q^{(n)}$ in momentum space are found using the properties in \refeqs{propertiespos}{propertiesmom}. After performing all Fourier integrations, we can write the quark propagators $S_\psi^{(n)}$ and $\tilde{S}_\psi^{(n)}$ as well as the gluon field strength as
\begin{align}
    S^{(n)}_\psi(p) &=  (ig_s)^n S^{(0)}_\psi(p) \left( \prod_{m=1}^n \gamma^{\mu_m}  A_{\mu_m}^{a_m} (k_m) t_{a_m} S^{(0)}_\psi ( p - \sum_{l=1}^m k_l ) \right) \ ,
    \\
    \tilde S^{(n)}_\psi (p) &=  (ig_s)^n \left( \prod_{m=1}^n S^{(0)}_\psi ( p + \sum_{l=1}^m \tilde k_l ) \gamma^{\mu_m}  A_{\mu_m}^{a_m} (\tilde k_m) t_{a_m} \right)  S^{(0)}_\psi(p) \ ,
    \\
    A_\mu^a (k,z_0) &= \sum_{n=0}^{\infty}\frac{1}{n!(n+2)} G_{\mu_0\mu;\mu_1\cdots\mu_n}^a\prod_{m=0}^n \left( i\frac{\partial}{\partial k_{\mu_m}} - z_0^{\mu_m} \right) \Big|_{k_{\mu_m}=0}.
\end{align}
Note that the products above are matrix products in colour space, so their ordering is relevant.
With the convention of \refeq{corrposition}, the free propagator is
\begin{equation}
    S_\psi^{(0)}(p) = \tilde{S}_\psi^{(0)}(p) = \frac{i(\slashed p+m_\psi)}{p^2-m_\psi^2}.
\end{equation}
The calculation is further simplified by the identity
\begin{equation}\label{eq:usefulder}
    \begin{aligned}
        \frac{\partial}{\partial k_\mu} \Big|_{k=0} S^{(0)}_\psi (p \pm k) = \pm i S^{(0)}_\psi (p) \gamma^\mu S^{(0)}_\psi (p).
    \end{aligned}
\end{equation}

To cross-check the calculation, we implemented two independent codes to derive the final analytic results.
In the first implementation, the results were obtained in \texttt{Mathematica}, using \texttt{Package-X 2.0}~\cite{Patel:2016fam} for the Dirac traces and loop integrals.
In the second implementation, the traces in both Dirac and colour space were carried out in \texttt{Mathematica} with \texttt{FeynCalc}.
The resulting loop integrals were then simplified using a combination of \texttt{FeynCalc}~\cite{Shtabovenko:2023idz}, \texttt{FeynHelpers}~\cite{Shtabovenko:2016whf}, and \texttt{LiteRed2}~\cite{Lee:2012cn}.
This code is included as an ancillary file attached to the arXiv version of this paper under the name \texttt{TensorCondensates.nb}.
Both implementations were carried out while retaining the full $z_0$ dependence.
The parameter $z_0$ cancels only after summing all diagrams in \reffig{GGcond} or, likewise, all diagrams in \reffig{GGGcond} as noted already in Ref.~\cite{Nikolaev:1982rq}.
By applying this procedure, we derive analytical results for all Wilson coefficients $\C_{\Gamma,4}^J$ and $\C_{\Gamma,6}^J$.
These Wilson coefficients are also provided as ancillary files in \texttt{Mathematica} format under the name $\mathtt{Cq2\_\Gamma\_J\_cond}$.
At dimension four one has $\mathtt{cond}=\mathtt{GG}$, whereas at dimension six one has $\mathtt{cond}=\mathtt{GGG}$ or $\mathtt{cond}=\mathtt{jj}$ (cf. \refeqs{OPEexpl}{C6}).

The analytic expressions for the Wilson coefficients $\C_{\Gamma,4}^J$ and $\C_{\Gamma,6}^J$ are considerably involved, containing square roots and logarithmic terms.
Performing the Borel transform of these expressions directly would be an extremely difficult, if not prohibitive, task.
Therefore, we employ the following strategy.
We expand $\C_{\Gamma,4}^J$ and $\C_{\Gamma,6}^J$ in the limit $m_c/m_b \to 0$, retaining terms up to $\order{m_c^4/m_b^4}$.
Higher-order contributions of $\order{m_c^5/m_b^5}$ are of order $10^{-3}$ and can thus be safely neglected, as they represent sub-per-mille corrections to already subleading terms.
The resulting expressions can then be Borel transformed.
Using the ordinary definition of the Borel transform (see, e.g., Ref.~\cite{Colangelo:2000dp}), one can derive the following expressions:
\begin{equation}
\begin{aligned}
    \mathcal{B}_{M^2}\left[(q^2)^n\right] & = 0 
    \,,\\
    \mathcal{B}_{M^2}\left[\frac{1}{(s-q^2)^n}\right]
    & =
    \frac{1}{(n-1)!}\,\frac{e^{-\frac{s}{M^2}}}{M^{2(n-1)}} 
    \,,\\
    \mathcal{B}_{M^2} \left[\frac{\log{(s-q^2)}} {(s-q^2)^{n}}\right] &= \frac{1}{(n-1)!}\frac{e^{-\frac{s}{M^2}}}{M^{2(n-1)}} \left[\log{(M^2)}+\Psi(n)\right]\ ,
\end{aligned}
\end{equation}
where $\Psi(n)$ is the digamma function and $n>0$. 
We also provide the Borel-transformed results as ancillary files in machine-readable Mathematica format under the name $\mathtt{CM2\_\Gamma\_J\_cond}$.

\subsection{QCD sum rules for the decay constants}

At this point, deriving the QCD sum rules for the decay constants becomes straightforward.
First, we equate the hadronic and OPE representations of the correlator:
\begin{align}
    \label{eq:sumrule-master}
    \Pi_\Gamma^J(M^2)\Big|_\OPE
    =
    \frac{1}{\pi} 
    \int\limits_0^\infty ds \, e^{-s/M^2} \, \Im\Pi_\Gamma^J(s)\Big|_\had
    \,.
\end{align}
Then, following the standard QCD sum-rule approach~\cite{Shifman:1978bx,Colangelo:2000dp}, the contributions from excited and continuum states are approximated by the perturbative part of the OPE above an effective threshold~$\hat{s}_\Gamma^J$ (usually denoted with $s_0$):
\begin{align}
    \label{eq:duality}
    \int\limits_{s_\Gamma^J}^{\infty} ds\, e^{-s/M^2}\, \rho_\Gamma^J(s)
    \approx
    \frac{1}{\pi}
    \int\limits_{\hat{s}_\Gamma^J}^{\infty} ds\, e^{-s/M^2}\, 
    \Im\C_{\Gamma,0}^J(s)
    \,.
\end{align}
This assumption, known as \emph{semi-global quark--hadron duality}, effectively isolates the ground-state contribution, while modelling the higher part of the spectrum by the perturbative continuum.
Since the duality approximation in \refeq{duality} is imposed only on the perturbative contribution, the condensate terms are not written in dispersive form.
This avoids the threshold singularities that appear in the imaginary parts of the gluon-condensate Wilson coefficients in a dispersive representation; see, e.g., Refs.~\cite{Bagan:1994dy,Colangelo:2026gfq}.
Whether or not one also subtracts the continuum tail of the condensate contributions when applying quark--hadron duality is a matter of prescription.
In the present analysis this ambiguity is numerically irrelevant, because the condensate contributions are negligible and their tails are further suppressed by the higher powers of the denominators.

Substituting Eqs.~\eqref{eq:hadT}, \eqref{eq:OPEexpl}, \eqref{eq:pertdisp}, and~\eqref{eq:duality} into \refeq{sumrule-master} and rearranging the result, we obtain
\begin{align}
    \label{eq:final-sumrule-fT}
    m_{B_c^*}^{4} \left(f_{B_c^*}^T\right)^2 e^{-m_{B_c^*}^2/M^2}
    &=
    \frac{1}{\pi}
    \int\limits_0^{\hat{s}_T^1}
        ds\, e^{-s/M^2} \,
        \Im\C_{T,0}^1(s)
        + 
    \C_{T,4}^1(M^2) \langle \cO_4 \rangle
    + 
    \C_{T,6}^1(M^2) \langle \cO_6 \rangle 
    \,.
\end{align}
Equation~\eqref{eq:final-sumrule-fT} constitutes the central relation used to extract the decay constants.
The QCD sum rules for the remaining decay constants are derived following an analogous procedure and read 
\begin{align}
    m_{B_{c0}^*}^{2} \left(f_{B_{c0}^*}\right)^2 e^{-m_{B_{c0}^*}^2/M^2}
    & =
    \frac{1}{\pi}
    \!\int\limits_0^{\hat{s}_V^0} \!
        ds\, e^{-s/M^2} \,
        \Im\C_{V,0}^0(s)
    + 
    \C_{V,4}^0(M^2) \langle \cO_4 \rangle
    + 
    \C_{V,6}^0(M^2) \langle \cO_6 \rangle 
    \,,
    \label{eq:final-sumrule-fV0}
    \\
    m_{B_c}^{2} \left(f_{B_c}\right)^2 e^{-m_{B_c}^2/M^2}
    & =
    \frac{1}{\pi}
    \int\limits_0^{\hat{s}_A^0} 
        ds\, e^{-s/M^2} \,
        \Im\C_{A,0}^0(s)
    + 
    \C_{A,4}^0(M^2) \langle \cO_4 \rangle
    + 
    \C_{A,6}^0(M^2) \langle \cO_6 \rangle 
    \,,
    \\
    m_{B_c^*}^{2} \left(f_{B_c^*}\right)^2 e^{-m_{B_c^*}^2/M^2}
    & =
    \frac{1}{\pi}
    \int\limits_0^{\hat{s}_V^1} 
        ds\, e^{-s/M^2} \,
        \Im\C_{V,0}^1(s)
    + 
    \C_{V,4}^1(M^2) \langle \cO_4 \rangle
    + 
    \C_{V,6}^1(M^2) \langle \cO_6 \rangle 
    \,,
    \\
    m_{B_{c1}}^{2} \left(f_{B_{c1}}\right)^2 e^{-m_{B_{c1}}^2/M^2}
    & =
    \frac{1}{\pi}
    \int\limits_0^{\hat{s}_{A}^1} 
        ds\, e^{-s/M^2} \,
        \Im\C_{A,0}^1(s)
    + 
    \C_{A,4}^1(M^2) \langle \cO_4 \rangle
    + 
    \C_{A,6}^1(M^2) \langle \cO_6 \rangle 
    \,,
    \label{eq:final-sumrule-fA1}
    \\
    m_{B_{c1}}^{4} \left(f_{B_{c1}}^T\right)^2 e^{-m_{B_{c1}}^2/M^2}
    & =
    \frac{1}{\pi}
    \int\limits_0^{\hat{s}_{AT}^1} 
        ds\, e^{-s/M^2} \,
        \Im\C_{AT,0}^1(s)
    + 
    \C_{AT,4}^1(M^2) \langle \cO_4 \rangle
    + 
    \C_{AT,6}^1(M^2) \langle \cO_6 \rangle 
    \,.
    \label{eq:final-sumrule-fAT}
\end{align}

The effective continuum threshold $\hat{s}_\Gamma^J$ is a key parameter in QCD sum-rule analyses.
Rather than fixing it by hand, we determine $\hat{s}_\Gamma^J$ dynamically using the \emph{daughter sum rule} (also known as the mass sum rule).
For instance, taking the logarithmic derivative of \refeq{final-sumrule-fT} with respect to $1/M^2$ yields
\begin{align}
    \label{eq:daughter}
    m_{B_c^*}^2 
    =
    \frac{
        \displaystyle
        \frac{1}{\pi} 
        \int\limits_0^{\hat{s}_T^1} ds \, s \, e^{-s/M^2} \, \Im\C_{T,0}^1(s)
        + 
        \frac{d}{d(-1/M^2)}\Big[
            \C_{T,4}^1(M^2) \langle \cO_4 \rangle
            + 
            \C_{T,6}^1(M^2) \langle \cO_6 \rangle
        \Big]
    }{
        \displaystyle
        \frac{1}{\pi} 
        \int\limits_0^{\hat{s}_T^1} ds \, e^{-s/M^2} \, \Im\C_{T,0}^1(s)
        + 
        \C_{T,4}^1(M^2) \langle \cO_4 \rangle
        + 
        \C_{T,6}^1(M^2) \langle \cO_6 \rangle
    }
    \,.
\end{align}
For a given value of $M^2$, the threshold $\hat{s}_T^1$ is fixed by requiring the right-hand side of \refeq{daughter} to approximate as closely as possible the physical meson mass squared $m_{B_c^*}^2$ (see \reftab{spectr}) without necessarily reproducing it exactly.
This procedure ensures maximal consistency between the OPE and hadronic sides of the sum rule.
The daughter sum rules corresponding to \refeqs{final-sumrule-fV0}{final-sumrule-fAT} can be derived straightforwardly.

Once $\hat{s}_\Gamma^J$ has been fixed by the daughter sum rules, the decay constants follow from \refeqs{final-sumrule-fT}{final-sumrule-fAT}.
The only parameter that remains to be discussed is the Borel parameter $M^2$, for which an appropriate window must be chosen.
We choose the working Borel window $[M^2_{\rm min}, M^2_{\rm max}]$ by balancing two competing requirements:
(\emph{i}) an adequate suppression of contributions from higher-dimensional condensates, and
(\emph{ii}) the dominance of the dispersive integral over the region $[0,\hat{s}_\Gamma^J]$ relative to the remaining tail of that integral.
The numerical analysis of these stability conditions and the final results for all decay constants are presented in the next section.

\section{Numerical analysis and results}
\label{sec:results}

Having derived the QCD sum rules, we now perform their numerical analysis to determine the central values and uncertainties of the decay constants.
The uncertainties of the meson masses listed in \reftab{spectr} are at the per mille level and are therefore negligible compared to the other sources of uncertainty.
The remaining input parameters are listed in \reftab{inputs} and are discussed below.
For the quark masses, we adopt the \MSbar scheme, which provides a short-distance mass definition and avoids the infrared sensitivities inherent in the on-shell scheme.
Similarly to the meson masses, the uncertainties associated with the quark masses and with $\alpha_s$ at their reference scales are negligible.
For the two-gluon condensate, we adopt the conservative estimate of Ref.~\cite{Ioffe:2002ee}, as obtained from QCD sum-rule analyses of charmonium decays. 
The value of the three-gluon condensate is obtained through the instanton model approximation (see Ref.~\cite{Shifman:1978bx}) using the values given in Ref.~\cite{Ioffe:2002ee}.  
Following Ref.~\cite{Shifman:1978bx}, we estimate the four-quark condensate using the vacuum-saturation, or factorisation, approximation. Possible deviations from this approximation are parametrised by $r_{vac}$, such that
\begin{align}
    \bra{0} g_s^4 j_\mu^a j_\mu^a \ket{0} 
    = 
    -r_{vac} \frac{16}{9} g_s^4 \bra{0} \bar u u + \bar d d +\bar ss \ket{0}^2 \,
    .
\end{align}

\begin{table}[t!]
    \newcommand{\pp}{\phantom{+}}
    \centering
    \renewcommand{\arraystretch}{1.4}
    \begin{tabular}{ccc}
        \toprule
        Parameter                                                                      &
        Cen. val.$\pm$unc./Interval                                                    &
        Ref.                                                                           \\
        \toprule
        $\overline{m}_b(\overline{m}_b)$                                               & 
        $4.183\pm0.004\,\GeV$                                                          & 
        \cite{PDG:2024cfk}                                                             \\
        $\overline{m}_c(\overline{m}_c)$                                               & 
        $1.273\pm0.003\,\GeV$                                                          & 
        \cite{PDG:2024cfk}                                                             \\
        $\alpha_s(m_Z)$                                                                &
        $0.1180\pm 0.0009$                                                             &
        \cite{PDG:2024cfk}                                                             \\
        $ \bra{0}\frac{\alpha_s}{\pi} G^a_{\mu\nu} G^a_{\mu\nu} \ket{0}$               &               
        $0.009\pm 0.007\,\GeV^4$                                                       & 
        \cite{Ioffe:2002ee}                                                            \\
        $\bra{0} g_s^3 f^{abc} G^a_{\mu\nu} G^b_{\nu\lambda} G^c_{\lambda\mu} \ket{0} $  &               
        $0.13\pm 0.10\,\GeV^6$                                                         & 
        \cite{Shifman:1978bx,Ioffe:2002ee}                                             \\
        $\langle \bar{u}u \rangle \equiv \langle \bar{d}d \rangle$                     &
        $-(0.278 \pm 0.022\ \GeV)^3$                                                   &
        \cite{FlavourLatticeAveragingGroupFLAG:2021npn}                                \\
        $\langle \bar{s}s \rangle /\langle \bar{u}u \rangle$                           &
        $0.8 \pm 0.3$                                                                  &
        \cite{Ioffe:2002ee}                                                            \\
        $r_{vac}$                                                                      &
        $1.0 \pm 0.5$                                                                  &
        \cite{Ioffe:2002ee}                                                            \\
        $M^2$                                                                          & 
        $[6,9]\,\GeV^2$                                                                & 
        ---                                                                            \\
        $\mu$                                                                          & 
        $[3,9]\,\GeV$                                                                  & 
        ---                                                                            \\
        \bottomrule
    \end{tabular}
    \caption{
    \label{tab:inputs} 
      Inputs used for the numerical evaluation of the QCD sum rules.
      The meson masses can be found in \reftab{spectr}.
      For the value of $r_{vac}$, we follow Ref.~\cite{Ioffe:2002ee}, but adopt a somewhat more conservative fifty-percent uncertainty on the deviation from the vacuum insertion approximation.
    }
\end{table}

\begin{figure}[p]
    \centering
    \resizebox{\linewidth}{!}{
            \input{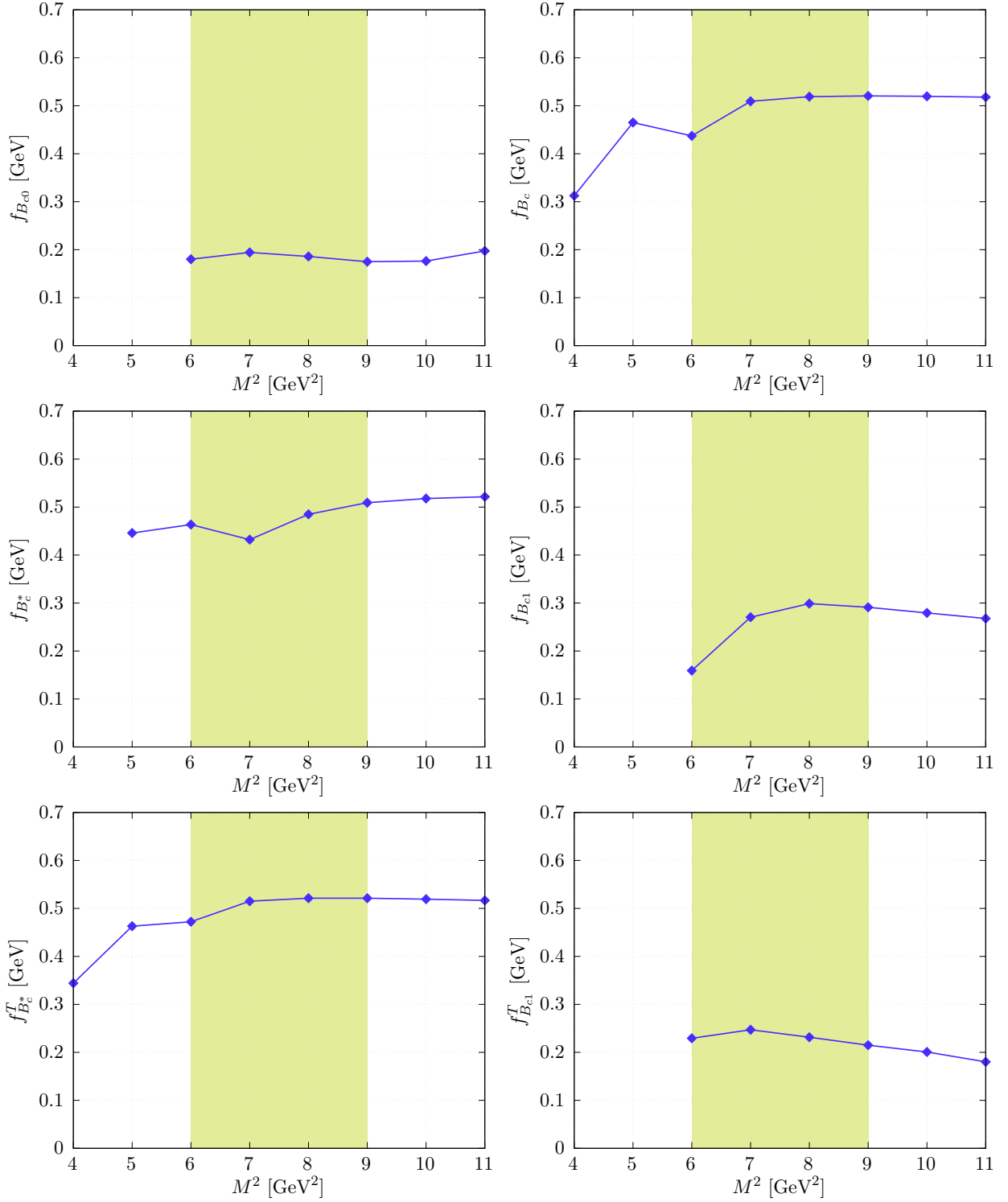}}
    \captionsetup{format=plain} 
    \caption{
        Borel stability of the decay constants extracted from the six QCD sum rules.
        Each panel shows the result obtained after fixing the effective continuum threshold from the corresponding daughter sum rule, as a function of the Borel parameter $M^2$ and for the representative renormalisation scale $\mu=6\,\GeV$.
        The shaded band indicates the working window $M^2\in[6,9]\,\GeV^2$ used in the final analysis.
        Points that fail the pole-contribution or mass-reproduction cuts are not shown.
    }
    \label{fig:decayconstants}
\end{figure}

We identify the interval $M^2 \in [6,9]\,\GeV^2$ as a suitable Borel window, selected according to the criteria outlined at the end of the previous section.
As shown in \reffig{decayconstants}, the decay constants are relatively stable throughout this range.
We vary the renormalisation scale of the quark masses and $\alpha_s$ over the range \hbox{$\mu \in [3,9]\,\text{GeV}$}.
This interval is intentionally broader than those adopted in, for example, Refs.~\cite{Gelhausen:2013wia,Pullin:2021ebn}, and therefore leads to a more conservative estimate of the uncertainty.
The quark masses and $\alpha_s$ are evolved to the required scale using four-loop renormalisation-group running as implemented in \texttt{RunDec}~\cite{Chetyrkin:2000yt,Herren:2017osy}.
The tensor current has a non-vanishing anomalous dimension, and the corresponding tensor decay constants are therefore scale dependent.
We do not vary the scale associated with this anomalous dimension, but fix it to $\mu_b \equiv \overline{m}_b(\overline{m}_b)$.

In practice, we discretise the $M^2$ and $\mu$ intervals into 49 and 19 equally spaced points, respectively.
This results in a total of 931 points for each sum rule in \refeqs{final-sumrule-fT}{final-sumrule-fAT}.
For every point in each sum rule, we determine the effective continuum threshold $\hat{s}_\Gamma^J$ according to the procedure described in the previous section.

Despite the general stability of the results within the $M^2$ and $\mu$ intervals we obtain some unphysical imaginary values for the decay constants. 
To avoid the appearance of these unphysical results and use the sum rules where they are more reliable we impose conditions on the $M^2$ and $\mu$ points.
To this end, we introduce two parameters: the pole contribution (PC) and the relative deviation on the meson mass obtained from the daughter sum rule, denoted by $\varepsilon_m$.
They are defined as
\begin{align}
    \text{PC}
    &:=
    \frac{
        \int\limits_{0}^{\hat{s}_\Gamma^J} ds\, e^{-s/M^2}\,
        \Im\C_{\Gamma,0}^J(s)
    }{
        \int\limits_{0}^{\infty} ds\, e^{-s/M^2}\,
        \Im\C_{\Gamma,0}^J(s)
    }\,,
    \\
    \label{eq:epsm}
    \varepsilon_m
    &:=
    \frac{|m^{\text{SR}}-m|}{m}
    \,.
\end{align}
Here, $m$ denotes the mass of one of the mesons listed in \reftab{spectr}, while $m^{\text{SR}}$ is the corresponding value obtained from the daughter sum rule (cf. \refeq{daughter}).
We only retain points satisfying $\text{PC}>50\%$ and $\varepsilon_m < 25\%$.
We observe a posteriori that these cuts remove all unphysical values of the decay constants and lead to an effective suppression of the condensate and NLO contributions.
For all sum rules except for two, almost all points pass the cuts.
The only exceptions are the two sum rules involving the $B_{c1}$ state, namely those in \refeqa{final-sumrule-fA1}{final-sumrule-fAT}, for which about $60\%$ of the points still pass the cuts.
More details are given in \refapp{Numerics}.

As expected, the values of the effective continuum threshold $\hat{s}_\Gamma^J$ lie slightly above the squared mass of the meson under consideration, with a standard deviation of a few $\GeV^2$.
We find
\begin{align}
\begin{aligned}
    \hat{s}_V^0 
    =&\, 52\pm4\,\text{GeV}^2 \,,&\qquad
    \hat{s}_V^1 
    =&\, 46\pm4\,\text{GeV}^2 \,,&\qquad
    \hat{s}_T^1 
    =&\, 49\pm3\,\text{GeV}^2 \,,\\
    \hat{s}_A^0 
    =&\, 49\pm4\,\text{GeV}^2 \,,&
    \hat{s}_A^1 
    =&\, 56\pm4\,\text{GeV}^2 \,,&
    \hat{s}_{AT}^1 
    =&\, 54\pm4\,\text{GeV}^2 \,.
\end{aligned}
\end{align}
These values are in substantial agreement with the thresholds used in previous QCD sum rule calculations~\cite{Colangelo:1992cx,Wang:2012kw,Aliev:2019wcm,Narison:2019tym,Narison:2020wql,Wang:2024fwc}.

\begin{table}[t!]
    \newcommand{\pp}{\phantom{+}}
    \centering
    \renewcommand{\arraystretch}{1.4}
    \begin{tabular}{cc}
        \toprule
        Decay constant                       &
        Cen. val.$\pm$unc. [$\MeV$]          \\
        \midrule
        $f_{B_{c0}^*}$                       &
        $193\pm 29$                          \\
        $f_{B_c}$                            &
        $499\pm 71$                          \\
        $f_{B_c^*}$                          &
        $480\pm 49$                          \\
        $f_{B_{c1}}$                         &
        $286\pm 47$                          \\
        $f_{B_c^*}^T$                        &
        $507\pm 65$                          \\
        $f_{B_{c1}}^T$                       &
        $244\pm 42$                          \\
        \bottomrule
    \end{tabular}
    \caption{
    \label{tab:results}
    Central values and uncertainties of the $B_c$-meson decay constants obtained from our QCD sum rules.} 
\end{table}

The decay-constant values reported in \reftab{results} are obtained by taking, for each decay constant, the interval spanned by the minimum and maximum values found over the set of considered points in $M^2$ and $\mu$.
Accordingly, the quoted central values are defined as the midpoints of these intervals, $(f_{\min}+f_{\max})/2$, while the associated uncertainties are taken to be half their widths, $(f_{\max}-f_{\min})/2$.
Since $\hat{s}_\Gamma^J$ is redetermined at each point of the $(M^2,\mu)$ grid, the dependence of the decay constants on $M^2$, $\mu$, and the effective continuum threshold is correlated in our procedure.
A decomposition of the quoted uncertainty into separate components would require an artificial variation scheme, such as varying $\mu$ while holding $M^2$ and $\hat{s}_\Gamma^J$ fixed, which would not reproduce our analysis and could underestimate the combined uncertainty.
We therefore quote the envelope of the full correlated scan as a single conservative uncertainty.

We find that the numerical contribution of the dimension-four terms, i.e.\ the two-gluon condensate, is at least one order of magnitude smaller than the corresponding perturbative contribution.
The dimension-six terms are even more suppressed, being at least one order of magnitude smaller than the dimension-four contributions.
This pattern indicates a good numerical convergence of the non-perturbative part of the OPE, in agreement with what is found in previous works~\cite{Colangelo:1992cx,Wang:2012kw,Aliev:2019wcm,Narison:2019tym,Narison:2020wql,Wang:2024fwc}.
Among the dimension-six terms, it is worth noting that the dominant contribution arises from the four-quark condensate term, while the three-gluon condensate term is comparatively smaller.
To the best of our understanding, the four-quark condensate contribution was not included in Refs.~\cite{Narison:2019tym,Narison:2020wql}.

As a cross-check, we also extract the decay constants using subtracted dispersion relations instead of the Borel transform.
In this implementation, the ratio $Q^2/k$, where $k$ denotes the number of subtractions, plays the role of the Borel parameter and is varied within the same interval used for $M^2$ above.
At very low values of $k$, the PC is small, reflecting the poor suppression of the continuum tail.
Starting from $k\simeq 6$, the tail is sufficiently suppressed and the extracted decay constants become stable.
The resulting values are in close agreement with those in \reftab{results}.
\\

\begin{figure}[t!]
    \centering
    \includegraphics[width=\linewidth]{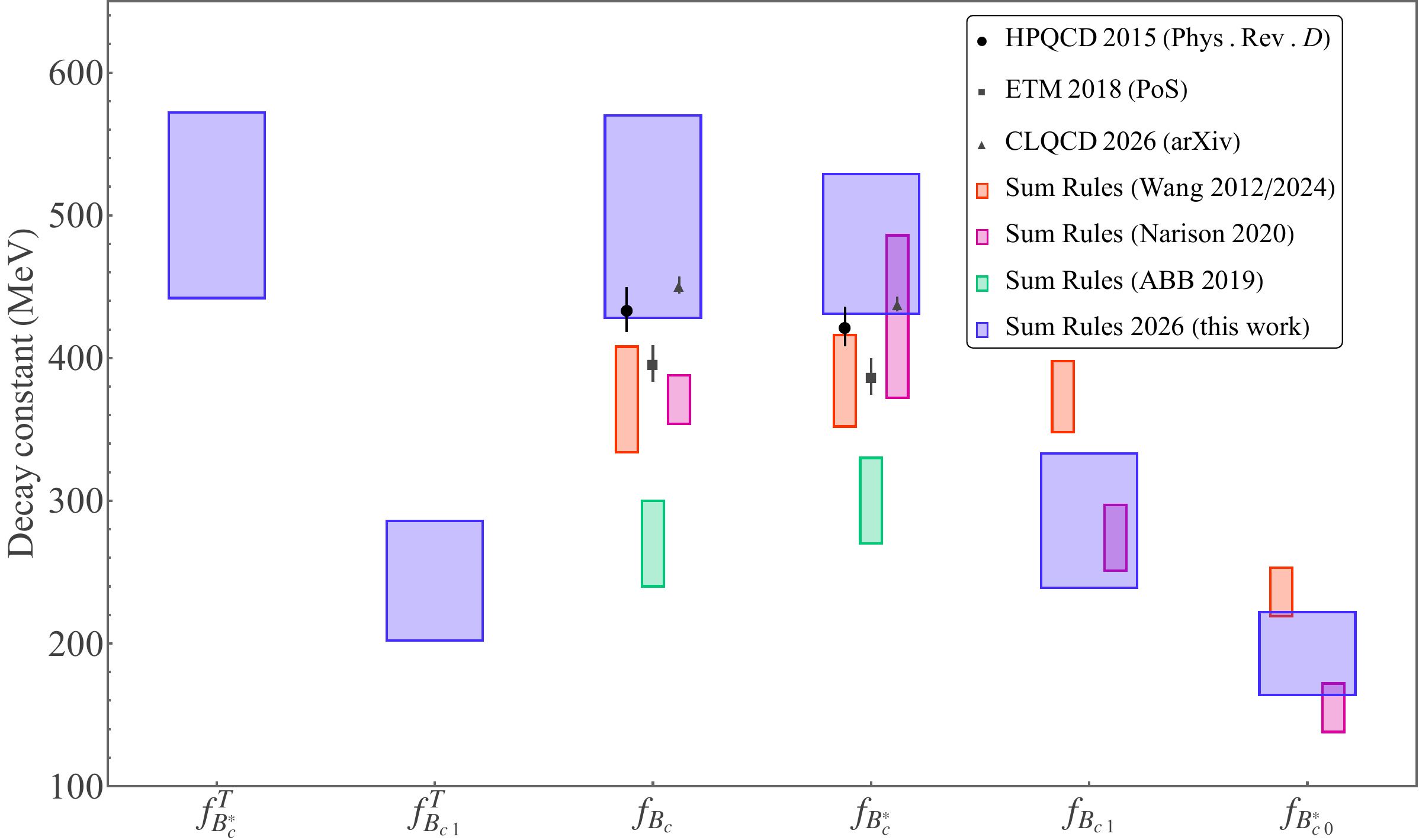}
    \caption{Comparison of our QCD sum rule predictions for all six $B_c$ decay constants with available lattice and sum rule determinations.
    Despite the current spread among lattice results of Refs.~\cite{Becirevic:2018qlo, Wingate:2021ycr,Cai:2026xja}, our predictions are consistent with the published lattice calculation of Ref.~\cite{Colquhoun:2015oha} within uncertainties.
    We discuss the treatment of uncertainties in our results versus other sum rule approaches of Refs.~\cite{Wang:2012kw,Wang:2024fwc,Narison:2020wql,Aliev:2019wcm} below.
    }
     \label{fig:latComp}
\end{figure}

In \reffig{latComp}, we compare our results for the six decay constants of the $B_c$ mesons with previous lattice QCD and QCD sum-rule determinations.
Although the currently available lattice QCD determinations appear to be in significant tension, we emphasise that only the result of Ref.~\cite{Colquhoun:2015oha} has been published in a peer-reviewed journal.
In fact, the ETM result reported in Ref.~\cite{Becirevic:2018qlo} appeared in the PoS proceedings and, as noted in Ref.~\cite{Wingate:2021ycr}, has not yet been finalised.
More recently, the CLQCD collaboration has reported a very precise result in a preprint~\cite{Cai:2026xja}, which is incompatible with the ETM determination.
For this reason, both the 2018 ETM result --- which lies about $2\sigma$ below the HPQCD determination~\cite{Colquhoun:2015oha} --- and the 2026 CLQCD result may still be subject to revision.
While the intrinsic uncertainties of the sum-rule approach prevent our results from attaining the level of precision currently achievable in lattice QCD, their agreement with the published lattice QCD determinations provides an important cross-check of our calculation.
In addition, we provide here, for the first time, determinations of the tensor-current decay constants of the $B_c^*$ and $B_{c1}$ mesons, for which no lattice QCD or sum-rule results are currently available.

\reffig{latComp} further shows that, although our sum rule results are consistent with the previous determinations of Refs.~\cite{Wang:2012kw,Narison:2020wql,Aliev:2019wcm,Wang:2024fwc}, our uncertainties are somewhat larger.
While the individual analyses differ in their truncation order in $\alpha_s$ and in the OPE, we attribute the main difference to the treatment of the renormalisation scale.
Indeed, previous works either fix the scale or vary it only over a very narrow range, whereas we adopt \hbox{$\mu \in [3,9]\,\text{GeV}$} in order not to underestimate the uncertainty associated with missing higher-order corrections in $\alpha_s$.
\\

As mentioned in the \hyperref[sec:intro]{Introduction}, the unitarity bound for the $b \to c$ form factors takes the form (cf. Ref.~\cite{Gubernari:2026sqc})
\begin{align}
    \label{eq:pre-Bound}
    \chi_\Gamma^J(Q^2)\Big|_{\OPE}
    =
    \chi_\Gamma^J(Q^2)\Big|_{\rm 1pt}
    +
    \chi_\Gamma^J(Q^2)\Big|_{\rm 2pt}
    +
    \dots
    \,,
\end{align}
where $\chi_\Gamma^J\big|_{\OPE}$ is a subtracted two-point correlator calculated in the framework of the OPE, in the spirit of \refsubsec{OPEcal} but without relying on semi-global quark--hadron duality.
The subtraction point $Q^2:=-q^2$ is required to satisfy $Q^2 \gg - (m_b + m_c)^2$ and is usually chosen as $Q^2 = 0$.
The terms $\chi_\Gamma^J\big|_{\rm 1pt}$ and $\chi_\Gamma^J\big|_{\rm 2pt}$ represent the contributions from one- and two-particle states, respectively, whereas the ellipsis stands for contributions from states of higher multiplicity.
The quantity $\chi_\Gamma^J\big|_{\rm 2pt}$ contains the contributions of the hadronic form factors entering the processes $B\to D^{(*)}$, $B_s\to D_s^{(*)}$, and $\Lambda_b\to \Lambda_c$, among others.
Assuming the minimal number of subtractions in  the corresponding dispersion relation, the one-particle contributions of the lowest-lying $B_c$ mesons in the relevant spin-parity channels take the form (cf. Ref.~\cite{Gubernari:2026sqc})
\begin{equation}
\begin{aligned}
    \label{eq:1pt-analyic}
    \chi_V^0(Q^2) \Big|_{\rm 1pt}
    &= 
    \frac{m_{B_{c0}^*}^2 f_{B_{c0}^*}^2}{(m_{B_{c0}^*}^2+Q^2)^{2}}
    + \dots \,,
    &
    \chi_A^0(Q^2) \Big|_{\rm 1pt}
    &= 
    \frac{m_{B_c}^2 f_{B_c}^2}{(m_{B_c}^2+Q^2)^{2}}
    + \dots \,,
    \\
    \chi_V^1(Q^2) \Big|_{\rm 1pt}
    &= 
    \frac{m_{B_c^*}^2 f_{B_c^*}^2}{(m_{B_c^*}^2+Q^2)^{3}}
    + \dots \,,
    &
    \chi_A^1(Q^2) \Big|_{\rm 1pt}
    &= 
    \frac{m_{B_{c1}}^2 f_{B_{c1}}^2}{(m_{B_{c1}}^2+Q^2)^{3}}
    + \dots \,,
    \\
    \chi_T^1(Q^2) \Big|_{\rm 1pt}
    &= 
    \frac{m_{B_c^*}^4 (f_{B_c^*}^T)^2}{(m_{B_c^*}^2+Q^2)^{4}}
    + \dots \,,
    &
    \chi_{AT}^1(Q^2) \Big|_{\rm 1pt}
    &= 
    \frac{m_{B_{c1}}^4 (f_{B_{c1}}^T)^2}{(m_{B_{c1}}^2+Q^2)^{4}}
    + \dots \,.
\end{aligned}    
\end{equation}
Here, the ellipsis denotes the contributions of other excited one-particle states.
As noted in Ref.~\cite{Caprini:1997mu} (see also the discussion in Ref.~\cite{Gubernari:2026sqc}), double counting is avoided by including only one-particle states with masses below that of the lightest multiparticle state considered in a given unitarity analysis.

Substituting the results of \reftab{results} into \refeq{1pt-analyic} and setting $Q^2=0$, we obtain
\begin{equation}
\begin{aligned}
    \label{eq:1pt-numeric}
    \chi_V^0(0) \Big|_{\rm 1pt}
    &= 6.05 \times 10^{-4}
    \,,
    &
    \chi_A^0(0) \Big|_{\rm 1pt}
    &= 4.65 \times 10^{-3}
    \,,
    \\
    \chi_V^1(0) \Big|_{\rm 1pt}
    &= 1.13 \times 10^{-4} \,\GeV^{-2}
    \,,
    &
    \chi_A^1(0) \Big|_{\rm 1pt}
    &= 2.79 \times 10^{-5} \,\GeV^{-2}
    \,,
    \\
    \chi_T^1(0) \Big|_{\rm 1pt}
    &= 1.21 \times 10^{-4} \,\GeV^{-2}
    \,,
    &
    \chi_{AT}^1(0) \Big|_{\rm 1pt}
    &= 1.98 \times 10^{-5} \,\GeV^{-2}
    \,.
\end{aligned}
\end{equation}
The numerical values quoted above are obtained by taking the lower endpoints of the ranges given in \reftab{results} and by neglecting the contributions of additional excited one-particle states.
This choice is conservative, since smaller one-particle contributions lead to a smaller saturation of the corresponding unitarity bound.
It is also instructive to compare the size of these contributions with the corresponding OPE results reported in Ref.~\cite{Gubernari:2026sqc}, namely the OPE susceptibilities entering the unitarity bounds.
The underlying correlator calculations were carried out in Refs.~\cite{Grigo:2012ji,Generet:2025hsv}.
We obtain
\begin{equation}
\begin{aligned}
    \frac{
        \chi_V^0(0) \Big|_{\rm 1pt}
    }{
        \chi_V^0(0) \Big|_\OPE
    }
    &= 9.40 \times 10^{-2}
    \,,
    &
    \frac{
        \chi_A^0(0) \Big|_{\rm 1pt}
    }{
        \chi_A^0(0) \Big|_\OPE
    }
    &= 1.89 \times 10^{-1}
    \,,
    \\
    \frac{
        \chi_V^1(0) \Big|_{\rm 1pt}
    }{
        \chi_V^1(0) \Big|_\OPE
    }
    &= 1.75 \times 10^{-1}
    \,,
    &
    \frac{
        \chi_A^1(0) \Big|_{\rm 1pt}
    }{
        \chi_A^1(0) \Big|_\OPE
    }
    &= 7.14 \times 10^{-2}
    \,,
    \\
    \frac{
        \chi_T^1(0) \Big|_{\rm 1pt}
    }{
        \chi_T^1(0) \Big|_\OPE
    }
    &= 2.49 \times 10^{-1}
    \,,
    &
    \frac{
        \chi_{AT}^1(0) \Big|_{\rm 1pt}
    }{
        \chi_{AT}^1(0) \Big|_\OPE
    }
    &= 8.04 \times 10^{-2}
    \,.
\end{aligned}
\end{equation}
These ratios quantify the fraction of the corresponding unitarity bounds saturated by the lowest one-particle states. 
It is well known that one-particle contributions play an important role in $b \to c$ unitarity analyses~\cite{Caprini:1997mu,Bigi:2017jbd,Bordone:2025jur}.
Indeed, as the values quoted above show, some one-particle contribution already saturate about $10\%$ of the unitarity bounds.
A particularly striking case is provided by the newly determined tensor decay constant of the $B_c^*$, which saturates about $25\%$ of the bound even under the conservative assumptions adopted in deriving this result.
Such sizeable saturations further constrain the form factor parameters in $b \to c$ transitions and can therefore improve the precision of observables in semileptonic decays, including the golden channels $\bar{B} \to D \ell \bar\nu$ and $\bar{B} \to D^* \ell \bar\nu$.
It should be noted, however, that in the modified BGL parametrisation of derived in Refs.~\cite{Gopal:2024mgb,Gubernari:2026sqc} the effect of these one-particle contributions is significantly more suppressed and therefore less relevant.
This is a consequence of the modifications to the dispersion relation underlying that parametrisation.
A quantitative assessment of the impact of these  one-particle contributions on the form factors, however, requires a dedicated analysis. 
In particular, it depends on the set of form-factor inputs used, their precision and correlations, their kinematic coverage, and the way in which the unitarity constraints are implemented.
This lies beyond the scope of the present work; see, e.g., Refs.~\cite{Martinelli:2023fwm,Gubernari:2026sqc} for dedicated analyses.

\section{Summary and conclusions}
\label{sec:summary}

We have presented a QCD sum-rule determination of the decay constants of the lowest-lying $B_c$ mesons in the $0^-$, $0^+$, $1^-$, and $1^+$ spin--parity channels.
The analysis includes the standard \mbox{(axial-)vector} currents, as well as the (axial-)tensor currents relevant for tensor interactions beyond the Standard Model.

The OPE used in this work includes the perturbative contribution at NLO in $\alpha_s$, together with the dimension-four and dimension-six corrections; namely the two-gluon, three-gluon, and light four-quark current-current condensate contributions.
For the tensor and axial-tensor currents, the Wilson coefficients of the dimension-four and dimension-six contributions have been derived here for the first time.
We provide the resulting Wilson coefficients, both before and after the Borel transformation, as machine-readable \texttt{Mathematica} ancillary files.
We also include the notebook \texttt{TensorCondensates.nb}, which contains an implementation of the calculation of these contributions.
The effective continuum thresholds have been fixed through the corresponding daughter sum rules, and the uncertainties have been estimated by varying the input parameters, the Borel parameter, and the renormalisation scale over conservative ranges.
The numerical hierarchy of the OPE contributions is well behaved: the dimension-four corrections are at least one order of magnitude smaller than the perturbative contribution, while the dimension-six corrections are further suppressed.

Our final results are collected in \reftab{results}, with the tensor-current decay constants quoted at $\mu_b=\overline{m}_b(\overline{m}_b)$.
The decay constants associated with (axial-)vector currents are compatible with previous QCD sum-rule calculations; for the $B_c$ and $B_c^*$ mesons, they are also compatible with the published lattice-QCD results.
The larger errors quoted here mostly reflect our conservative treatment of the renormalisation-scale dependence.
As an additional check, we extract the decay constants using subtracted dispersion relations instead of the Borel transform and find compatible results.

Finally, we have evaluated the corresponding one-particle contributions entering the unitarity bounds for $b\to c$ form factors.
These contributions are non-negligible in all channels and are particularly relevant in the tensor channel, where the lowest-lying $B_c^*$ state gives the largest saturation among the channels considered.
The decay constants obtained in this work therefore provide updated input for future dispersive analyses of $\bar{B}\to D^{(*)}$, $\bar{B}_s\to D_s^{(*)}$, and related semileptonic transitions, both in the Standard Model and in scenarios with tensor interactions.

\section*{Acknowledgements}

We sincerely thank A.~Khodjamirian for his valuable insights and discussions.
We are also grateful to Dr.~Narison for the correspondence and for providing information on the calculation presented in Ref.~\cite{Bagan:1994dy}.

This work has been funded by the Deutsche Forschungsgemeinschaft (DFG, German Research Foundation) -- Emmy-Noether Grant
No. 558599025.
\sloppy This work has also been partially supported by the STFC consolidated grants ST/T000694/1 and ST/X000664/1.

\appendix
\addcontentsline{toc}{section}{Appendices}

\newcounter{APP}
\renewcommand{\theAPP}{\Alph{APP}}
\setcounter{APP}{0}
\renewcommand{\theequation}{\theAPP.\arabic{equation}}
\renewcommand{\thesubsection}{\theAPP.\arabic{subsection}}

\appsection{Derivation of Vacuum-Averaged Condensates}
\label{app:Condensates}

The GG, GGG, GDDG, and DGDG condensates have been introduced in \refsubsec{OPEcal}. Their tensor decompositions are derived by considering the most general Lorentz structures antisymmetric in the indices of the gluon field strength tensors, and then properly normalising the results.

\subsection{GG Condensate}
By considering the most general Lorentz structure that complies with the antisymmetry in $\alpha$-$\beta$ and $\mu$-$\nu$, we can write
\begin{equation}
    \bra{0} G_{\alpha\beta}^a G_{\mu\nu}^a  \ket{0}=A\, (g_{\alpha \mu} g_{\beta \nu} - g_{\alpha \nu} g_{\beta \mu})\,.
\end{equation}
In order to determine the constant of proportionality, one has to project both sides such that the contractions of the Lorentz indices on the gluon field strength tensors match.
This amounts to multiplying both sides by $g^{\beta\mu}g^{\alpha\nu}$, which eliminates the possibility of having the Levi-Civita symbol and yields
\begin{equation}
    \bra{0} G_{\alpha\beta}^a G^{a}_{\beta\alpha} \ket{0} = A\, g^{\beta\mu}g^{\alpha\nu}(g_{\alpha \mu} g_{\beta \nu} - g_{\alpha \nu} g_{\beta \mu})=-12A\,.
\end{equation}
Solving for $A$, 
\begin{equation}
    A=-\frac{1}{12} \bra{0} G_{\alpha\beta}^a G^{a}_{\beta\alpha} \ket{0}=\frac{1}{12} \bra{0} G_{\alpha\beta}^a G^{a}_{\alpha\beta} \ket{0}\,,
\end{equation}
one finally obtains
\begin{equation}
    \bra{0} G_{\alpha\beta}^a G_{\mu\nu}^a  \ket{0}=\frac{1}{12} \bra{0} G_{\gamma\delta}^a G^{a}_{\gamma\delta} \ket{0}\, (g_{\alpha \mu} g_{\beta \nu} - g_{\alpha \nu} g_{\beta \mu})\,.
\end{equation}

\subsection{GGG Condensate}
The procedure is analogous to the one for the GG condensate, although now there is antisymmetry in $\mu$-$\nu$, $\alpha$-$\beta$ and $\rho$-$\sigma$. Starting from
\begin{equation}
\begin{aligned}
    \bra{0}f^{abc} G_{\mu\nu}^a G_{\alpha\beta}^b G_{\rho\sigma}^c  \ket{0}&=B\, \left( g_{\mu\sigma}g_{\alpha\nu}g_{\beta\rho} 
    + g_{\mu\beta}g_{\alpha\rho}g_{\sigma\nu} 
    + g_{\alpha\sigma}g_{\mu\rho}g_{\nu\beta} 
    + g_{\rho\nu}g_{\mu\alpha}g_{\beta\sigma} \right. \\
    &- g_{\mu\beta}g_{\alpha\sigma}g_{\rho\nu} 
    - g_{\mu\sigma}g_{\alpha\rho}g_{\nu\beta} 
    - g_{\alpha\nu}g_{\mu\rho}g_{\beta\sigma} 
    - g_{\beta\rho}g_{\mu\alpha}g_{\nu\sigma} )\,,
\end{aligned}
\end{equation}
The coefficient $B$ is obtained by contracting both sides with $g^{\nu \alpha} g^{\beta \rho} g^{\mu \sigma}$,
\begin{equation}
    B=\frac{1}{24}\,\bra{0}f^{abc} G_{\mu\nu}^a G^{b}_{\nu\rho} G_{\rho\mu}^{c}  \ket{0}\,,
\end{equation}
and, straightforwardly, one finds the results already given in Refs.~\cite{Nikolaev:1982rq,Reinders:1984sr}:
\begin{equation}
\label{eq:gggcond}
\begin{aligned}
    \bra{0} f_{abc} G_{\mu\nu}^a G_{\alpha\beta}^b G_{\rho\sigma}^c \ket{0} 
    &= 
    \frac{1}{24} \bra{0} f_{abc} G_{\gamma\delta}^a G^{b}_{\delta\varepsilon} G_{\varepsilon\gamma}^c \ket{0}
    \\ 
    &\times ( g_{\mu\sigma}g_{\alpha\nu}g_{\beta\rho} 
    + g_{\mu\beta}g_{\alpha\rho}g_{\sigma\nu} 
    + g_{\alpha\sigma}g_{\mu\rho}g_{\nu\beta} 
    + g_{\rho\nu}g_{\mu\alpha}g_{\beta\sigma} \\
    &
    - g_{\mu\beta}g_{\alpha\sigma}g_{\rho\nu} 
    - g_{\mu\sigma}g_{\alpha\rho}g_{\nu\beta} 
    - g_{\alpha\nu}g_{\mu\rho}g_{\beta\sigma} 
    - g_{\beta\rho}g_{\mu\alpha}g_{\nu\sigma} )\ .
\end{aligned}
\end{equation}

\subsection{GDDG Condensate}

The derivation for this condensate is slightly more involved. We start by rewriting the expression using the definition of the commutator,
\begin{equation} \label{eq:gddg_Start}
    \begin{aligned}
        \bra{0} G_{\mu\nu}^a  G_{\alpha\beta;\rho\sigma}^a  \ket{0}
        &=\bra{0} G_{\mu\nu}^a (D_\sigma
        D_\rho)^{ab} G_{\alpha\beta}^b  \ket{0}
        = \bra{0} G_{\mu\nu}^a ( [D_\sigma,D_\rho]  + D_\rho 
        D_\sigma )^{ab}G_{\alpha\beta}^b \ket{0} \\
        &= -g_sf^{cab} \bra{0} G_{\mu\nu}^a G_{\sigma\rho}^c G_{\alpha\beta}^b \ket{0} 
        + \bra{0} G_{\mu\nu}^a  G_{\alpha\beta;\sigma\rho}^a  \ket{0}\,,
    \end{aligned}
\end{equation}
where we used $[D_\sigma,D_\rho]^{ab}=-ig_sG_{\sigma\rho}^c(\sigma^c)^{ab}=-g_sf^{cab}G_{\sigma\rho}^c$ and $\sigma^c$ is the $SU(3)_C$ generator in the adjoint representation.
Adding the LHS to both sides of the equation and dividing by two, we obtain

\begin{equation}
\label{eq:GDDGexpanded}
    \bra{0} G_{\mu\nu}^a  G_{\alpha\beta;\rho\sigma}^a  \ket{0}= -\frac{1}{2}g_sf^{cab} \bra{0} G_{\mu\nu}^a G_{\sigma\rho}^c G_{\alpha\beta}^b \ket{0} + \bra{0} G_{\mu\nu}^a G_{\alpha\beta;(\rho\sigma)}^a  \ket{0}\,,
\end{equation}
where we introduced the usual symmetrisation notation $X_{(\mu\nu)} = \frac{1}{2}(X_{\mu\nu} + X_{\nu\mu})$. We now look more closely at the Lorentz structure of the gluon condensates on the RHS of the above equation. Starting with the first term, we directly identify the GGG condensate of \refeq{gggcond}. Accounting for the antisymmetry of both the gluon field strength tensor and the structure constants,
\begin{equation}
     -\frac{1}{2}g_sf^{cab} \bra{0} G_{\mu\nu}^a G_{\sigma\rho}^c G_{\alpha\beta}^b \ket{0} = \frac{1}{2}g_s \bra{0} f^{abc}G_{\mu\nu}^a G_{\alpha\beta}^b G_{\rho\sigma}^c  \ket{0}\,.
\end{equation}
For the second term, following from the symmetries in the indices, we can write
\begin{align} \label{eq:gddglorentz}
    \bra{0} G_{\mu\nu}^a G_{\alpha\beta;(\rho\sigma)}^a  \ket{0}
    &=
    C_1 \, 
    g_{\rho \sigma} (g_{\mu \alpha} g_{\nu \beta} - g_{\mu \beta} g_{\nu \alpha}) \notag \\
    &+
    C_2 \,\big[
    -g_{\alpha  \sigma } g_{\beta  \nu } g_{\mu  \rho}
    +g_{\alpha  \sigma } g_{\beta  \mu } g_{\nu  \rho }
    +g_{\alpha \nu } g_{\beta  \sigma } g_{\mu  \rho}
    -g_{\alpha  \mu } g_{\beta  \sigma } g_{\nu  \rho } 
    \notag \\
    &
    -g_{\alpha  \rho } g_{\beta  \nu } g_{\mu  \sigma }
    +g_{\alpha  \nu } g_{\beta  \rho } g_{\mu  \sigma }
    +g_{\alpha \rho } g_{\beta  \mu } g_{\nu  \sigma }
    -g_{\alpha  \mu } g_{\beta  \rho } g_{\nu  \sigma}\ \big].
\end{align}
In this case, the constants of proportionality are obtained via a system of equations by contracting both sides with $g^{\nu\alpha}g^{\beta\mu}g^{\sigma\rho}$ and $g^{\mu\rho}g^{\nu\alpha}g^{\beta\sigma}$, 
obtaining, respectively,
\begin{align}\label{eq:gddgcontraction}
    \bra{0} G^{a}_{\mu\alpha} G_{\alpha\mu;(\sigma\sigma)}^a  \ket{0} 
    &=
    -48 C_1 + 48 C_2 \, ,
    \\
    \bra{0} G^{a}_{\mu\alpha} G_{\alpha\sigma;(\mu\sigma)}^a  \ket{0} 
    &=
    -12 C_1 + 48 C_2  \, ,
\end{align}
which implies
\begin{align}
    C_1 
    &
    = \frac{1}{36} \big(  
    -\, \bra{0} G^{a}_{\mu\alpha} G_{\alpha\mu;(\sigma\sigma)}^a  \ket{0} +
    \bra{0} G^{a}_{\mu\alpha} G_{\alpha\sigma;(\mu\sigma)}^a  \ket{0}
    \big) \,,
    \\
    C_2 
    &
    = \frac{1}{144} \big(  
    - \bra{0} G^{a}_{\mu\alpha} G_{\alpha\mu;(\sigma\sigma)}^a  \ket{0} +
    4\, \bra{0} G^{a}_{\mu\alpha} G_{\alpha\sigma;(\mu\sigma)}^a  \ket{0}
    \big) \, .
\end{align}
From here, we use the relations
\begin{align}
    G_{\mu\nu;\alpha\alpha}^{a} &= g_s\,(2 f^{abc} G_{\mu\alpha}^{b} G_{\nu \alpha}^c +j_{\mu;\nu}^a - j_{\nu;\mu}^a) \,, \\
    G^a_{\alpha \beta; \mu \nu} 
    &= g_s f^{abc} G^b_{\alpha \beta} G^c_{\mu \nu} + G^a_{\alpha \beta; \nu \mu} \, ,
    \\
    G^a_{\nu \mu;\mu} &= g_s j^a_\nu \, , \\
    \bra{0} j_{\mu;\nu}^a G^a_{\mu\nu} \ket{0} &= -g_s \bra{0}j_\mu^a j_\mu^a \ket{0} \,,
\end{align}
where $j_{\mu;\nu}^a := (D_\nu)^{ab} j^b_\mu$ in complete analogy with \refeq{GDnotation}.
These identities imply
\begin{align}\label{eq:GDDGtoGGG}
    \bra{0} G^{a}_{\mu\alpha} G_{\alpha\mu;(\sigma\sigma)}^a  \ket{0} 
    &=
    -2g_s \bra{0} f^{abc} G^a_{\mu \alpha} G^b_{\alpha \sigma} G^c_{\sigma \mu} \ket{0} +2 g_s^2 \bra{0} j^a_\mu j^a_\mu \ket{0} \, , \\
    \bra{0} G^{a}_{\mu\alpha} G_{\alpha\sigma;(\mu\sigma)}^a  \ket{0} 
    &=
    -\frac{1}{2}
    g_s \bra{0} f^{abc} G^a_{\mu \alpha} G^b_{\alpha \sigma} G^c_{\sigma \mu} \ket{0} +
    g_s^2 \bra{0} j^a_\mu j^a_\mu \ket{0} 
    \, ,
\end{align}
which leads to the solutions
\begin{align}
    C_1 &= \frac{1}{36} \big( \frac{3}{2} g_s  \bra{0} f^{abc} G^a_{\mu \alpha} G^b_{\alpha \sigma} G^c_{\sigma \mu}  \ket{0} -g_s^2 \bra{0} j^a_\mu j^a_\mu \ket{0}  \big) \, , \\
    C_2 &= \frac{1}{72} \big(  g_s^2 \bra{0} j^a_\mu j^a_\mu \ket{0}  \big) \, .
\end{align}
Putting the terms symmetric and anti-symmetric in $\rho$-$\sigma$ together and using \refeq{gggcond}, we finally arrive at
\begin{align}
    \bra{0} G_{\mu\nu}^a  G_{\alpha\beta;\rho\sigma}^a  \ket{0} &= \frac{g_s}{48} \bra{0} f_{abc} G_{\gamma\delta}^a G^{b}_{\delta\varepsilon} G_{\varepsilon\gamma}^{c} \ket{0}\notag
    \\* 
    &\times \big[\, 2 g_{\rho \sigma} (g_{\mu \alpha} g_{\nu \beta} - g_{\mu \beta} g_{\nu \alpha})\notag\\
    &+ g_{\mu\sigma}g_{\alpha\nu}g_{\beta\rho} 
    + g_{\mu\beta}g_{\alpha\rho}g_{\sigma\nu} 
    + g_{\alpha\sigma}g_{\mu\rho}g_{\nu\beta} 
    + g_{\rho\nu}g_{\mu\alpha}g_{\beta\sigma}\notag \\*
    &
    -\, g_{\mu\beta}g_{\alpha\sigma}g_{\rho\nu} 
    - g_{\mu\sigma}g_{\alpha\rho}g_{\nu\beta} 
    - g_{\alpha\nu}g_{\mu\rho}g_{\beta\sigma} 
    - g_{\beta\rho}g_{\mu\alpha}g_{\nu\sigma} \big]\,\notag \\
    &
    + \frac{g_s^2}{72} \bra{0} j_\lambda^a j_\lambda^a \ket{0} \notag \\
    &
    \times \big[
    2 g_{\rho  \sigma } \left(g_{\alpha  \nu } g_{\beta  \mu }
    -g_{\alpha  \mu } g_{\beta  \nu}\right) \notag \\
    &
    -g_{\alpha  \sigma } g_{\beta  \nu } g_{\mu  \rho}
    +g_{\alpha  \sigma } g_{\beta  \mu } g_{\nu  \rho }
    +g_{\alpha \nu } g_{\beta  \sigma } g_{\mu  \rho}
    -g_{\alpha  \mu } g_{\beta  \sigma } g_{\nu  \rho } 
    \notag \\
    &
    -g_{\alpha  \rho } g_{\beta  \nu } g_{\mu  \sigma }
    +g_{\alpha  \nu } g_{\beta  \rho } g_{\mu  \sigma }
    +g_{\alpha \rho } g_{\beta  \mu } g_{\nu  \sigma }
    -g_{\alpha  \mu } g_{\beta  \rho } g_{\nu  \sigma} \big]\,,
    \end{align}
which again agrees with Ref.~\cite{Nikolaev:1982rq}.

\subsection{DGDG Condensate}

Using the product rule, we can write
\begin{equation}
    \bra{0} G_{\mu \nu;\rho}^a G_{\alpha \beta; \sigma}^a \ket{0} = D_\rho \bra{0} G_{\mu \nu}^a G_{\alpha \beta; \sigma}^a \ket{0} - \bra{0} G_{\mu \nu}^a G_{\alpha \beta; \sigma\rho}^a \ket{0}\,.
\end{equation}
Since vacuum expectation values of total derivative operators vanish, we can effectively drop the first term on the RHS of the above equation.
The second term on the RHS of the above equation exactly matches the form of \refeq{gddg_Start}, but with $\rho$ and $\sigma$ swapped:
\begin{equation}
    \bra{0} G_{\mu \nu;\rho}^a G_{\alpha \beta; \sigma}^a \ket{0} = - \bra{0} G_{\mu\nu}^aG_{\alpha \beta; \sigma \rho}^a\ket{0}\,.
\end{equation}
This relation is implicit in previous literature; for completeness, we state it explicitly.

\appsection{Details on the numerical analysis}
\label{app:Numerics}

In this appendix, we collect additional information on the numerical stability checks used in the extraction of the decay constants.
For each channel and for each point of the $(M^2,\mu)$ grid, the effective continuum threshold is first obtained from the corresponding daughter sum rule.
The same point is then used in the original sum rule to extract the decay constant, and the point is retained only if it satisfies the pole-contribution and mass-reproduction cuts introduced in \refsec{results}.
The purpose of these cuts is not to tune the central value, but to discard regions in which the truncated OPE or the continuum approximation is visibly less reliable.
Consequently, the missing points in the low-$M^2$ part of the plots correspond to grid points that do not pass at least one of these cuts.

\begin{figure}[p]
    \centering

    \resizebox{\linewidth}{!}{
            \input{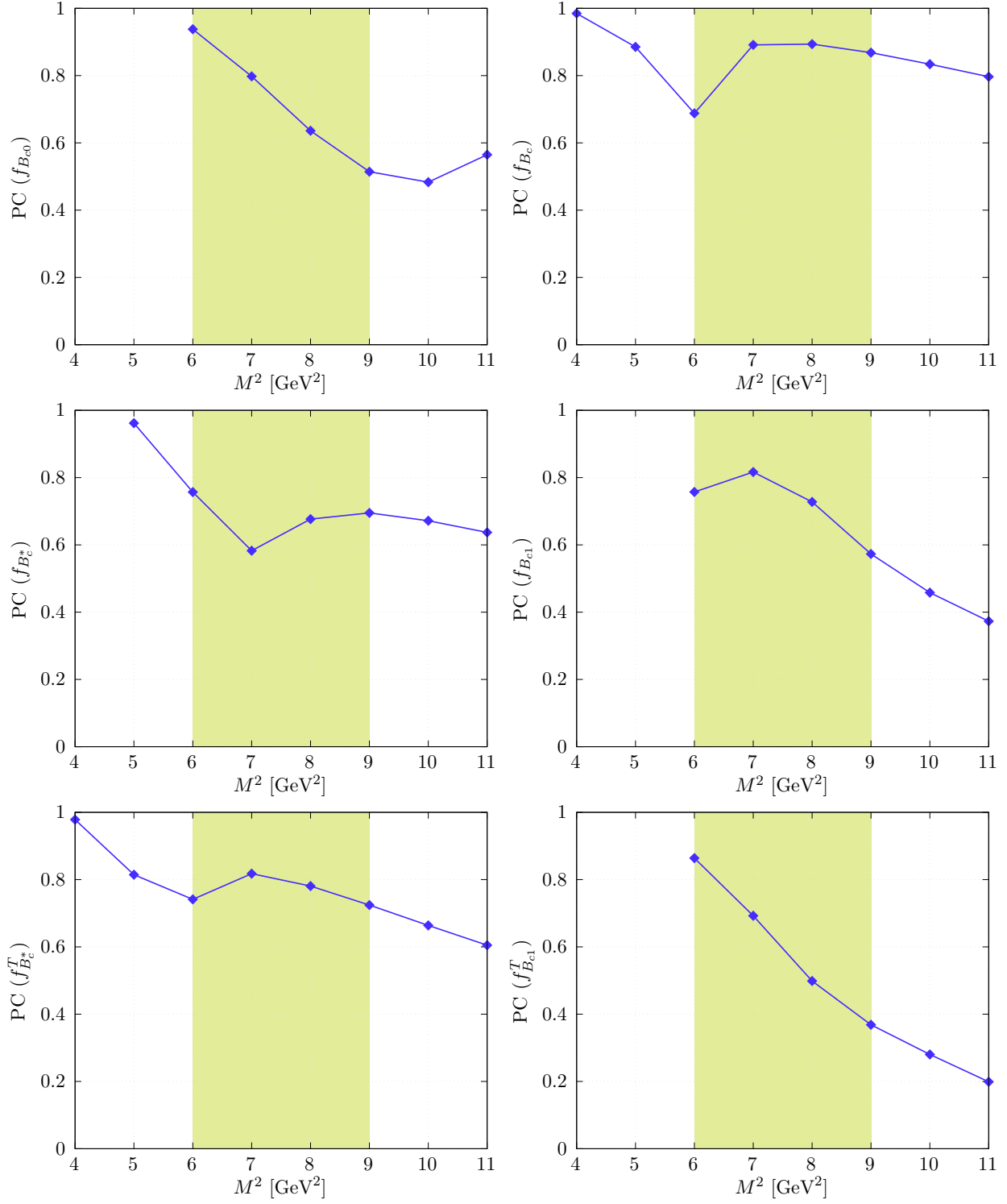}}
    \captionsetup{format=plain} 
    \caption{
        Pole contribution for the six QCD sum rules as a function of the Borel parameter $M^2$, evaluated at the representative renormalisation scale $\mu=6\,\GeV$.
        The shaded band indicates the working window $M^2\in[6,9]\,\GeV^2$ used to extract the decay constants.
    }
    \label{fig:pc}
\end{figure}

\reffig{pc} shows the pole contribution for the six correlators considered.
The pole contribution decreases as $M^2$ is increased, reflecting the reduced exponential suppression of the continuum at larger Borel masses.
The selected window $M^2\in[6,9]\,\GeV^2$ therefore represents a compromise: it keeps the continuum contribution under control while avoiding the low-$M^2$ region where the (daughter) sum rule is less stable and higher-dimensional condensates become more important.
Within this window the condition $\text{PC}>50\%$ is fulfilled for almost all points in the scalar, pseudoscalar, vector, and tensor channels.
The axial-vector and axial-tensor channels are the most restrictive ones and are responsible for most of the points rejected by this criterion.

\begin{figure}[p]
    \centering

    \resizebox{\linewidth}{!}{
            \input{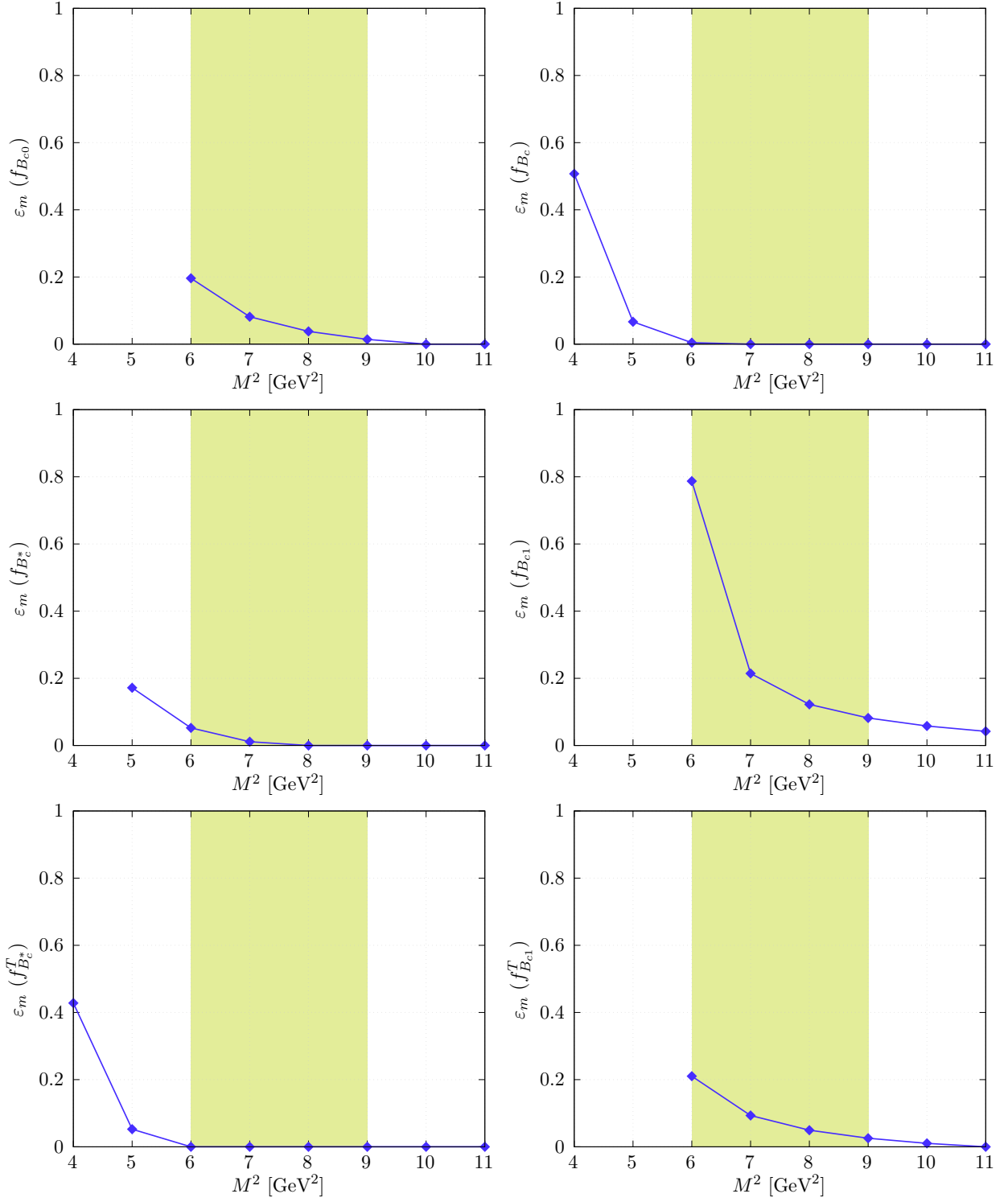}}
    \captionsetup{format=plain} 
    \caption{
        Relative mass deviation $\varepsilon_m$ obtained from the daughter sum rules in the six channels, evaluated at the representative renormalisation scale $\mu=6\,\GeV$.
        The shaded band indicates the working window $M^2\in[6,9]\,\GeV^2$ used in the final analysis.}
    \label{fig:epsilonm}
\end{figure}

\reffig{epsilonm} shows the corresponding relative deviation $\varepsilon_m$, as defined in \refeq{epsm}, obtained from the daughter sum rules.
This quantity measures how well the same effective threshold reproduces the physical meson mass before the decay constant is extracted.
For the retained points the deviations remain below the adopted $25\%$ tolerance, and in most channels they are considerably smaller.
The largest deviations again occur in the sum rules involving the $B_{c1}$ state, which explains why these channels have the smallest accepted fraction of grid points in the analysis.

Combining the two requirements removes the unphysical values encountered in the broader scan over $M^2$ and $\mu$.
The final intervals quoted in \reftab{results} are obtained from the envelope of the accepted points, so that the quoted uncertainties include the residual sensitivity to the Borel parameter, the renormalisation scale, and the effective continuum threshold.

\clearpage
\bibliographystyle{JHEP}
\bibliography{references}

\end{document}